\documentclass[a4paper,fleqn, review ]{cas-sc}
\usepackage{siunitx}
\usepackage{hyperref}
\usepackage[authoryear]{natbib}

\usepackage{amsmath} 
\usepackage{graphicx}
\usepackage{comment}
\usepackage{subcaption}
\usepackage{siunitx}
\DeclareSIUnit\ktonne{ktonnes}
\usepackage{float}
\usepackage{fancyhdr}
\usepackage{tabularx}
\usepackage{booktabs}
\usepackage{hyperref}

\def\tsc#1{\csdef{#1}{\textsc{\lowercase{#1}}\xspace}}
\tsc{WGM}
\tsc{QE}
\tsc{EP}
\tsc{PMS}
\tsc{BEC}
\tsc{DE}

\begin{document}
\let\WriteBookmarks\relax
\def\floatpagepagefraction{1}
\def\textpagefraction{.001}



\title [mode = title]{Validation of Dry Bulk Pile Volume
Estimation Algorithm based on Angle of
Repose using Experimental Images}                      

%
\author[1]{Madhu Koirala}[type=editor,                
                        prefix=,
                        role=,
                        orcid=0000-0001-7511-2910]

\cormark[1]

\ead{m.koirala@uit.no}

\affiliation[1]{organization={UiT The Arctic University of Norway},
    addressline={Lodve Langesgate 2}, 
    city={Narvik},
    postcode={8514}, 
    country={Norway}}

\author[1]{Pål Gunnar Ellingsen}

\author[2]{Ashenafi Zebene Woldaregay}[%
   role=,
   suffix=,
   ]

\affiliation[2]{organization={University Hospital of North Norway},
    addressline={Hansine Hansens veg 67}, 
    city={Tromsø},
    postcode={9019},     
    country={Norway}}

\cortext[cor1]{Corresponding author}

\begin{abstract}
Estimation of volume of piles in shipping ports plays a pivotal role for logistics management, facilitates better ship rescheduling and rerouting for economic benefits and contributes to overall  efficient shipping management. This paper presents validation results for a volume estimation algorithm for dry bulk cargo piles stored in open ports. Using remote sensing images obtained in a laboratory setting, the method first detects the contour of the pile and then reconstructs its 3D model based on the material's angle of repose, and estimates the volume accordingly. We validated the algorithm on full conical piles and single-ridge elongated piles, and further tested it on reclaimed conical and elongated piles. The results demonstrated the algorithm's strong potential for accurately estimating pile volume from experimental images and a reference satellite image, achieving high accuracy in our validation.

\end{abstract}



\maketitle

\section{Introduction}

Volume estimation of cargoes in shipping ports is very important for shipping companies to ensure efficient logistics management, stockpiles monitoring \citep{Monitoring_and_computation_of_the_volumes_of_stockpiles_of_bulk_material_by_means_of_UAV_photogrammetric_surveying, Stockpile_monitoring_using_linear_shape-from-shading_on_PlanetScope_imagery, Automatic_Stockpile_Volume_Monitoring_Using_Multi-View_Stereo_from_Skysat_Imagery, Vision_based_stockpile_inventory_measurement_using_uncrewed_aerial_systems} and better routing and scheduling of ships, especially tramp shipping, which is mainly driven by demand for cargo in a particular port \citep{The_value_of_foresight_in_the_drybulk_freight_market}. 

Traditionally, the calculation of the stockpile volume has been done using conventional survey methods such as total station (TS) \citep{A_new_approach_for_geo-monitoring_using_modern_total_stations_and_RGB+_D_images} and Global Navigation Satellite System (GNSS) \citep{Evaluation_study_of_GNSS_technology_and_traditional_surveying_in_DEM_generation_and_volumes_estimation}. These methods typically collect coordinate points on the surface of the object and create 3D model for volume estimation. Although these methods are accurate, they are very tedious and time consuming. Later, Light Detection and Ranging (LiDAR) \citep{Salt_stockpile_inventory_management_using_LiDAR_volumetric_measurements}, terrestrial laser scanning \citep{A_rapid_method_for_estimating_the_angle_of_repose_and_volume_of_grain_piles_using_terrestrial_laser_scanning}, and unmanned aerial vehicle (UAV) photogrammetry \citep{Monitoring_and_computation_of_the_volumes_of_stockpiles_of_bulk_material_by_means_of_UAV_photogrammetric_surveying} were introduced as more convenient and accurate methods of monitoring the stockpile volume. Methods similar to ours, where the researchers use satellite images for stockpile monitoring, are few and far between, but is slowly rising due to the availability of high resolution satellite images. \cite{Automatic_Stockpile_Volume_Monitoring_Using_Multi-View_Stereo_from_Skysat_Imagery} use SkySat imagery to calculate coal volume in Richard's Bay. The study corrects inconsistencies related to camera models, where camera models are represented as Rational Polynomial Cameras (RPCs). The refined RPCs are used to create different Digital Surface Models (DSMs) from stereo images. The DSMs are used to calculate the volume of stockpiles. Noisy or interpolated data in the photogrammetric DSMs used for volume estimation, however, can sometimes give inaccurate results. \cite{Estimating_construction_waste_truck_payload_volume_using_monocular_vision} use deep learning on a single image to detect the boundaries of buckets on truck to accurately quantify construction waste. The method requires prior knowledge of truck dimensions and range finders installed at the entrance of the storage facility. The study mentions that volume accuracy could be further increased by increasing the segmentation accuracy of the buckets. Linear Shape from Shading (SFS) was used to calculate the volume of coal piles from Capella SAR (radar) images by \cite{Interactive_Segmentation_for_Shape_From_Shading_Over_HR_SAR_Images}.  Before applying SFS, the authors make the boundaries of the piles more clear using an interactive segmentation method, which typically involves segmenting out the unnecessary parts and masking them, segmenting the piles, and refining the boundaries for better accuracy. This method is limited to materials with constant texture and reflectance properties. This also suffers from layover issues inherent in SAR imagery. SFS was used to calculate the volume of coal stocks in PlanetScope images by \cite{Stockpile_monitoring_using_linear_shape-from-shading_on_PlanetScope_imagery}. The method first delineates the area of interest (the region occupied by piles) and then applies SFS to generate 3D surface models, and is less accurate compared to UAV monitoring, but useful for overall stockpile monitoring. One of its drawbacks is that it does not solve the issue of occluded piles. The authors use high-resolution WorldView images to calculate the volume of a large tailings storage facility \citep{Satellite_monitoring_of_a_large_tailings_storage_facility}. Software from PhotoSat Ltd. is used to create Digital Elevation Models (DEMs) using satellite images. Elevation changes over time in these DEMs are used to calculate the volume of the tailings. 

In our previous work \citep{Identification_of_bulk_cargo_piles_stored_openly_at_ports_using_artificial_intelligence_and_their_volume_estimation_from_satellite_images}, we devised an algorithm based on the angle of repose and the inherent free-fall shape of a pile to estimate the volume of piles. Compared to traditional methods, which are more complex and time consuming, it estimates the volume of piles from the contour of piles using the angle of repose. Our previous work presented only a basic validation of the method using manually drawn contours of piles. In this paper, our objective is to perform a more rigorous validation through systematic laboratory experiments and a satellite image. Furthermore, we downsampled the original images to mimic lower resolution satellite images and tested the algorithm against these images. The use of satellite images for volume estimations makes it possible to monitor much broader area of interests and opens up new areas of applications such as enhancing tramp shipping efficiency, increasing global commodity flow visibility and enabling real-time flow monitoring.

\section{Materials and Methods}
The validation of the proposed algorithm is done using experimental images obtained under controlled laboratory conditions and a satellite image. Ideally, satellite images would have been used for this study, as they allow for the calculation of volume in ports of interest and support a wide range of applications. However, due to the difficulty in obtaining accurate ground truth volume data from actual ports, we conducted controlled experiments instead. In our experimental setup, we created various piles of different shapes and sizes, including conical, elongated, and reclaimed forms. Images of these piles were captured, their contours were extracted, and their volumes were calculated using the proposed algorithm.
\textcolor[rgb]{0.05,0.05,0.05}{\subsection{Image Pre-processing}
Although image preprocessing techniques like edge enhancement, contrast adjustment are often used to improve contour detection in the images, no preprocessing was applied in this study. Both experimental and satellite images showed sufficiently clear contours for accurate segmentation. However, in scenarios involving noisy or blurry contours, appropriate preprocessing may significantly improve contour detection and, consequently, volume estimation accuracy.}
\subsection{Experimental Setup}
\begin{figure}[h!bt]
    \centering
    \includegraphics[origin = c, width=0.3\columnwidth]{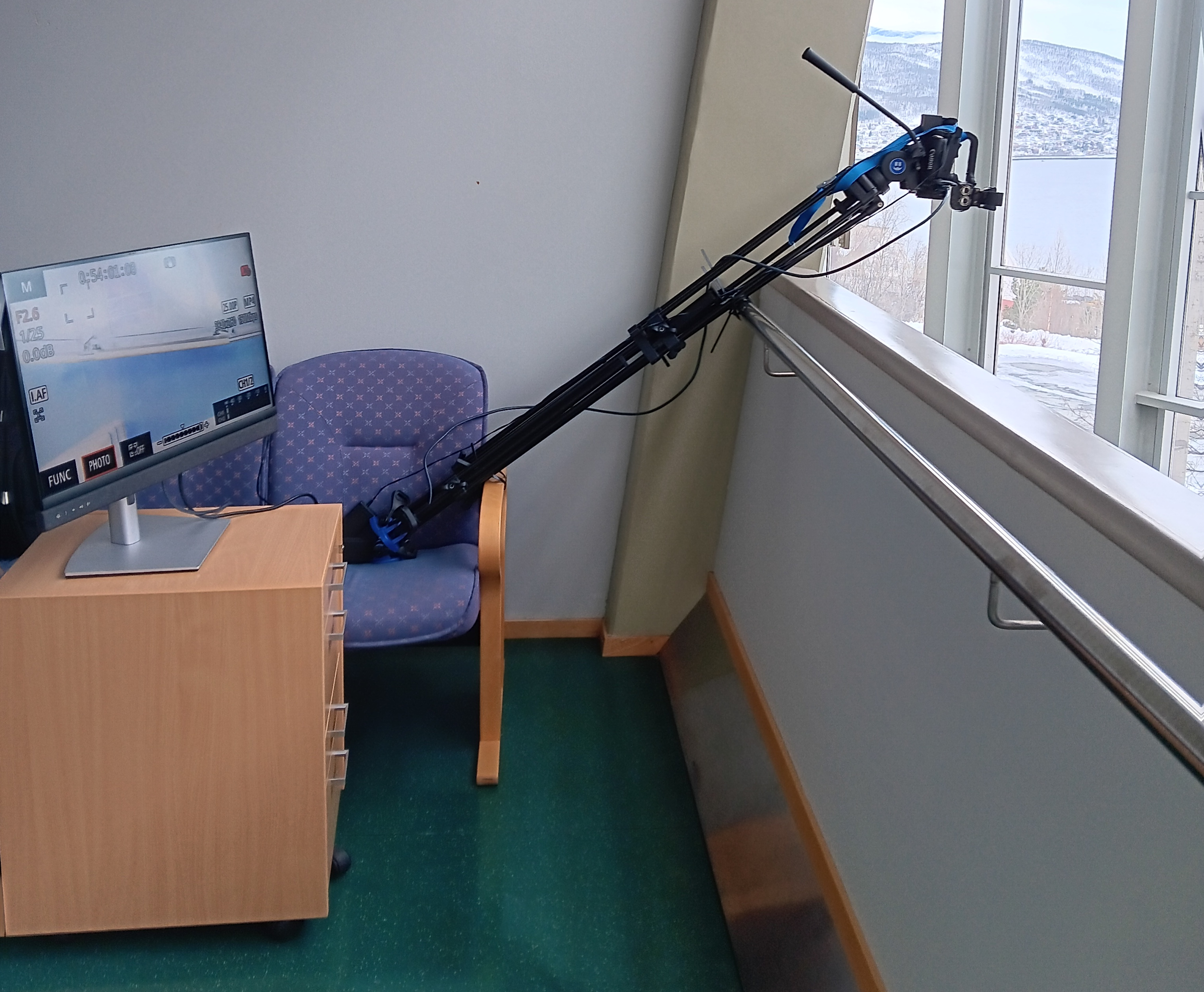}
    \includegraphics[origin = c, width=0.3\columnwidth]{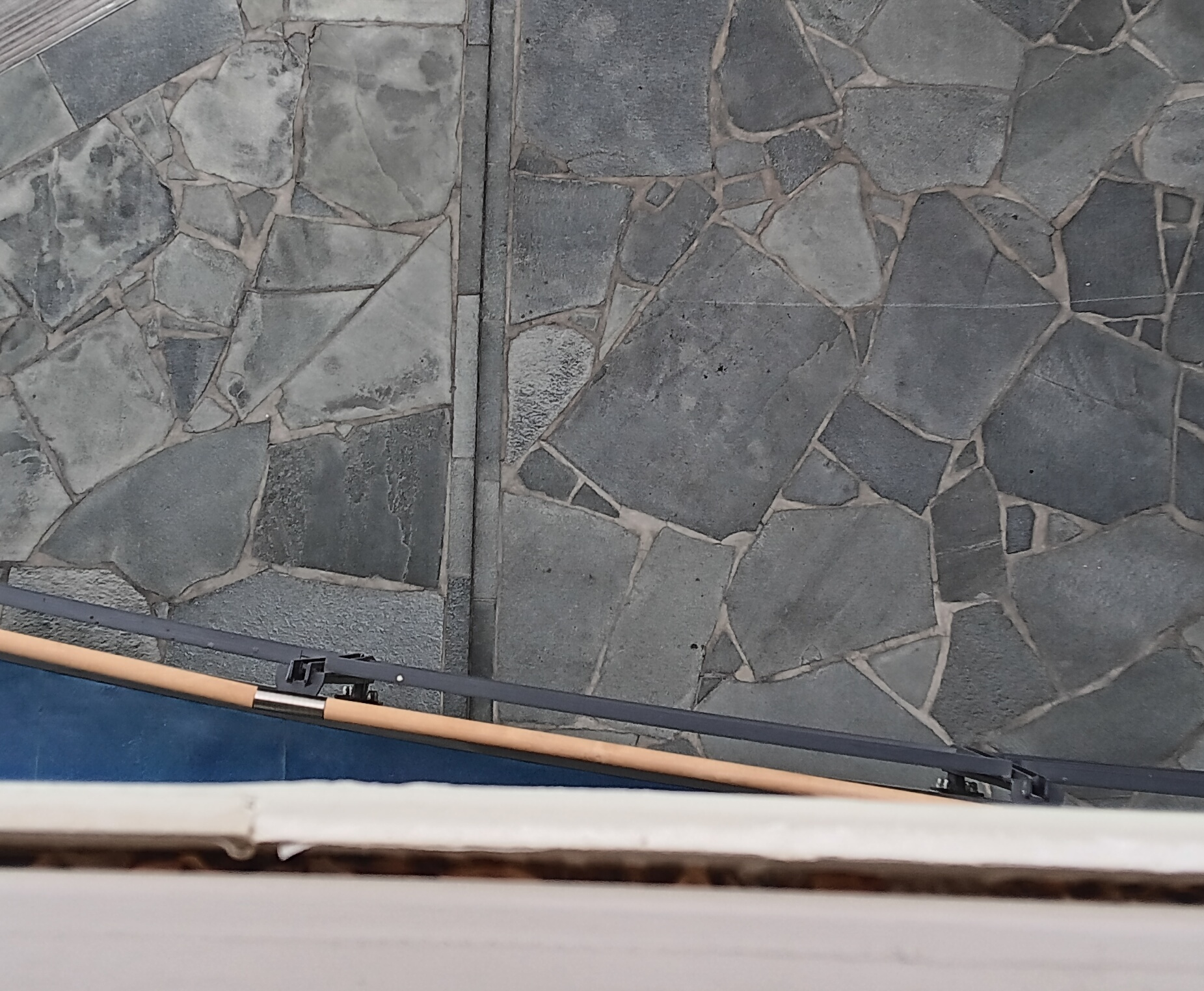}
    \includegraphics[origin = c, width=0.4\columnwidth]{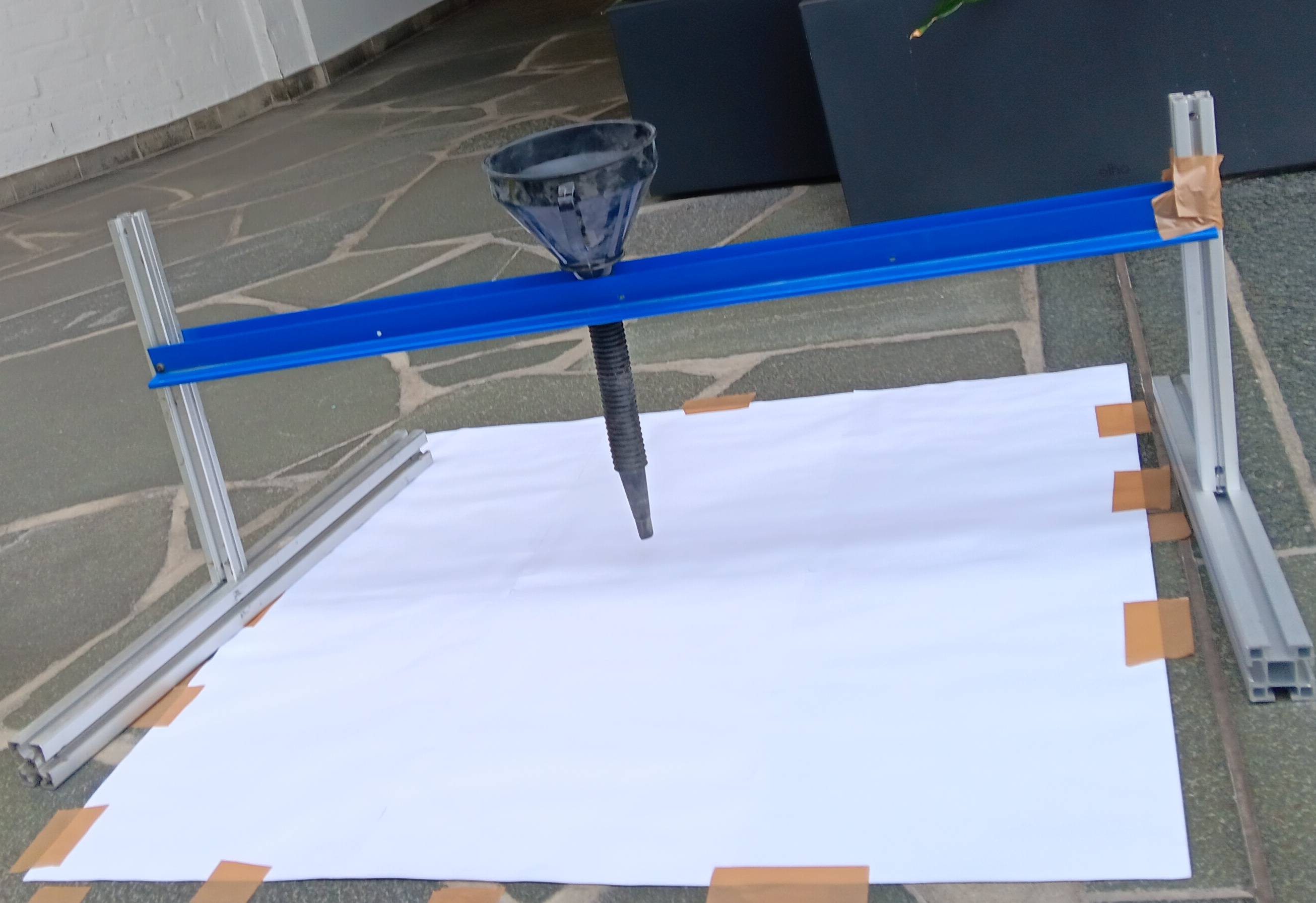}    
       
    \caption{Experimental Setup (The top-left shows camera setup on the third floor balcony, the top-right shows height of the camera from the area where piles are made, and the bottom shows pile-making stand and funnel along with the white sheet over which piles are made.)}
    \label{fig:experimental-setup}
\end{figure}

To validate the proposed volume estimation algorithm, we simulated satellite image capture using a digital camera. Given the difficulty of obtaining accurate ground truth volume data from an actual port, experimental images were used for validation. The experimental setup is illustrated in Figure~\ref{fig:experimental-setup}. A Canon XA40 4K camcorder was mounted vertically on a fixed stand at a height of \SI{8.4}{\m} above the ground, where the test piles were constructed. The height was measured using a digital laser distance meter with an accuracy of $\pm \SI{2}{mm}$. The entire process was monitored via an HDMI-connected display.

\begin{figure}[h!bt]
    \centering
    \includegraphics[origin = c, width=0.6\columnwidth]{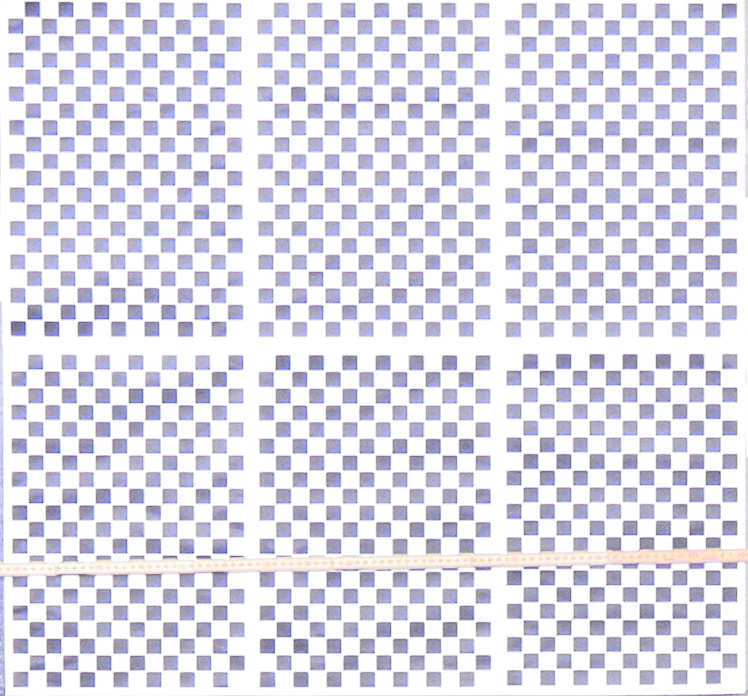}

    \caption{Checkboard (Grid of black and white Pixels of \SI{20}{mm} size) for verification of uniformity of pixel sizes in the image.}
    \label{fig:checkboard}
\end{figure}

For making the piles, we prepared two aluminum stands (white) in the T-shape, and one steel angle (blue) was placed on each side of the stands. These angles allow the funnel to slide back and forth to simulate the free fall of the material that makes a pile on the white sheet of paper on the ground. The angle was leveled on both stands using a spirit level to ensure it is perfectly parallel to the horizontal. We measured the distance of the entire image along its width and used this to convert the pixels into the real lengths of the piles. Only a small part near the center of the imaging area was used. The uniformity in the image was ensured by imaging and measuring a gridded paper with \SI{20}{\mm} by \SI{20}{\mm} white and black pixels as shown in Figure~\ref{fig:checkboard}. This gave us confirmation that lens correction was not needed. In addition, we made sure that there was no need for any geometric correction as the camera was rotated to align with the reference line used for measurement. Compared to our previous work~\citep{Geometric_Shape_and_Volume_Modelling_of_Dry_Bulk_Piles_using_a_Single_Image}, where we used sieved sand with particle diameters \SI{\leq2}{\mm}, we used playground sand with diameters of \SIrange{0.2}{1}{\mm}. To make a conical pile, we poured the sand through the funnel onto a single point on the ground until it reached a certain height using a measuring jug of maximum capacity of \SI{1}{\litre}. For elongated single-ridge piles, we first poured the conical pile to the required height, moving in small steps between each pour until the desired length. After making the piles, we simulated the reclaiming of the piles from the bottom of the piles. We removed sand from the bottom part of the piles using a small funnel.

\subsection{Segmenting the piles}
For pile segmentation, the online image annotation platform Computer Vision Annotation Tool ~\citep{cvat} was employed. The Segment Anything Model 2 (SAM-2)\citep{ravi2024sam2segmentimages}, integrated within CVAT, was used to automatically generate segmentation masks. Using this tool, segmentation could be initiated simply by clicking on the target piles, after which SAM-2 automatically created the corresponding masks. In cases where pile boundaries were unclear, manual adjustments were made to refine the segmentation results in addition to the automatic output provided by SAM-2.

\subsection{Volume Calculation}
After segmentation of the piles, the coordinates of the pile masks were obtained in JSON format. These coordinate values and angle of repose were used in the volume calculation algorithm to calculate the volume of each pile. We briefly present the algorithm proposed in our previous work \citep{Identification_of_bulk_cargo_piles_stored_openly_at_ports_using_artificial_intelligence_and_their_volume_estimation_from_satellite_images} as follows: (For a detailed explanation, refer to it.)

\begin{enumerate}
\item Detect piles in the image using the deep learning instance segmentation algorithm.
\item For each pile, iterate over all the points inside the contour of the pile to calculate the distance of the pixel from its nearest point in the contour.
\item For each pixel, use the angle of repose and its distance from the nearest point in the contour to calculate the height using a known angle of repose. 
\item Sum up all the height measurements. 
\item Use the resolution of the image (real dimension of each pixel) to convert the sum of height values into the actual volume estimation. 
\end{enumerate}

\section{Results}
This section shows all the experimental images that we have used for the validation of the volume algorithm. Furthermore, corresponding segmentation masks and reconstructed 3D piles are also presented here.
\begin{figure*}[!htb]
\centering
\includegraphics[width=0.16\columnwidth]{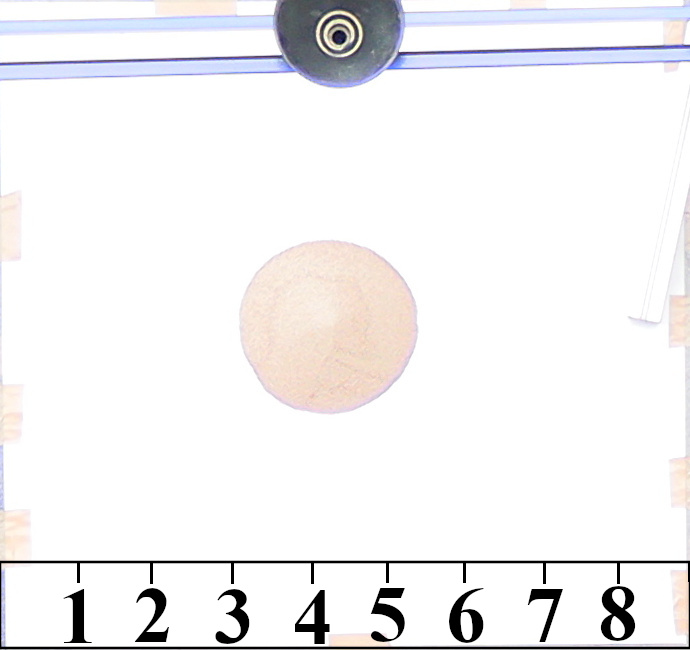}
\includegraphics[width=0.16\columnwidth]{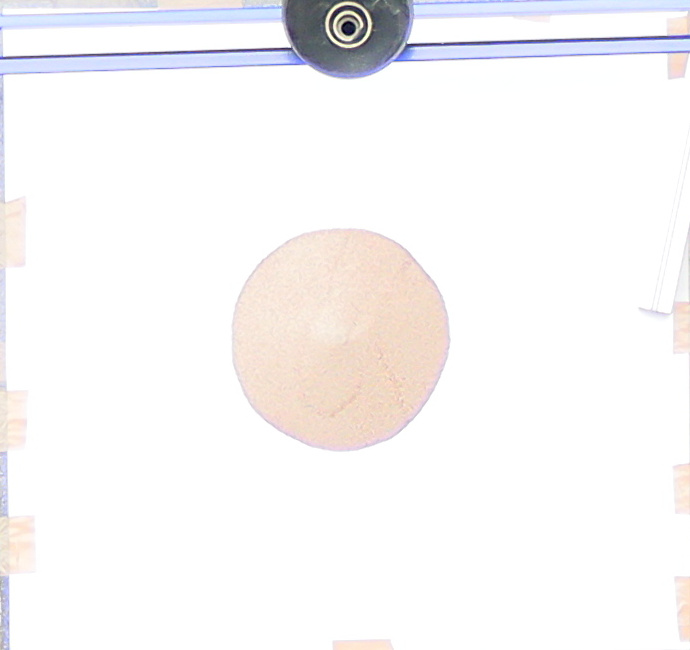}
\includegraphics[width=0.16\columnwidth]{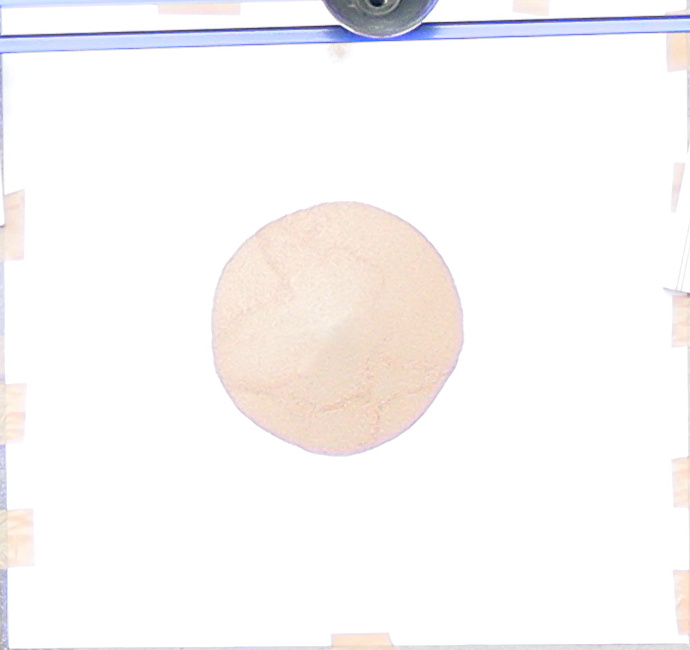}
\includegraphics[width=0.16\columnwidth]{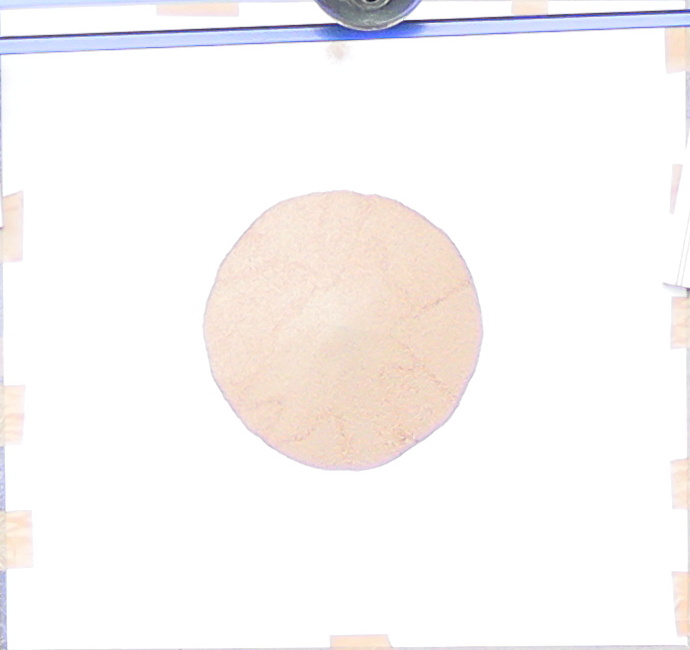}
\includegraphics[width=0.16\columnwidth]{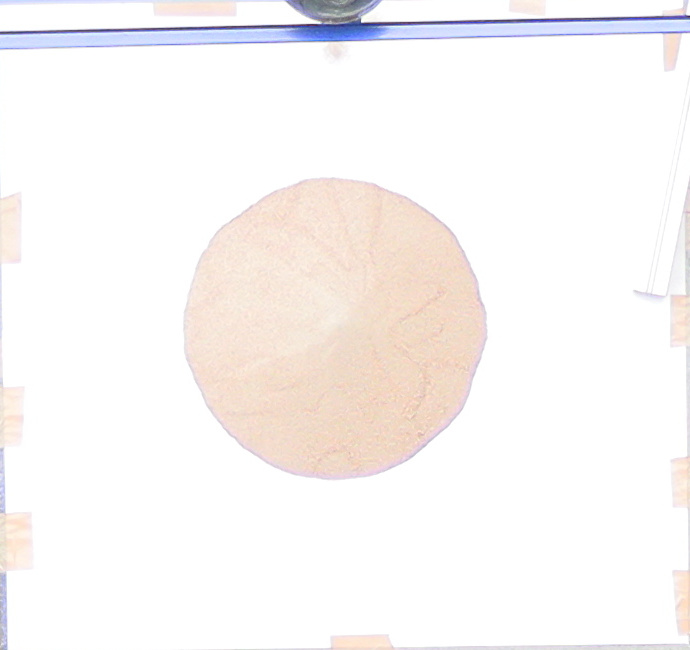}
\includegraphics[width=0.16\columnwidth]{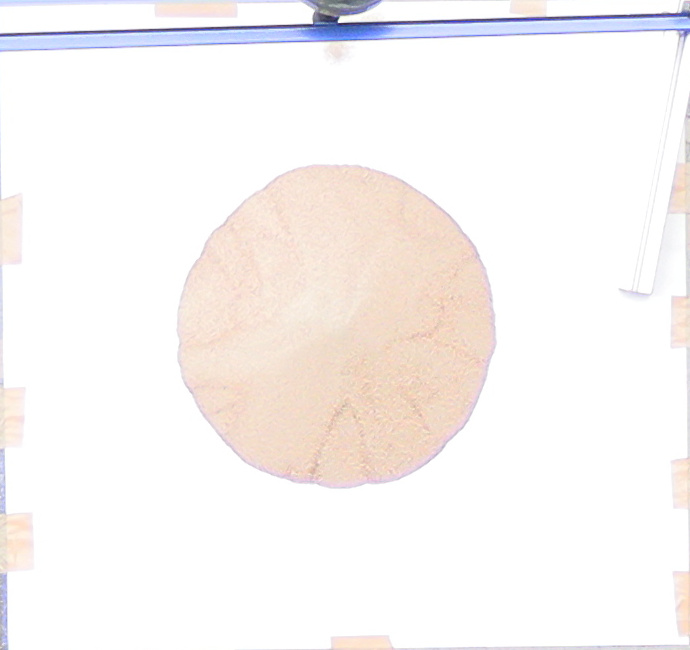}
\includegraphics[width=0.16\columnwidth]{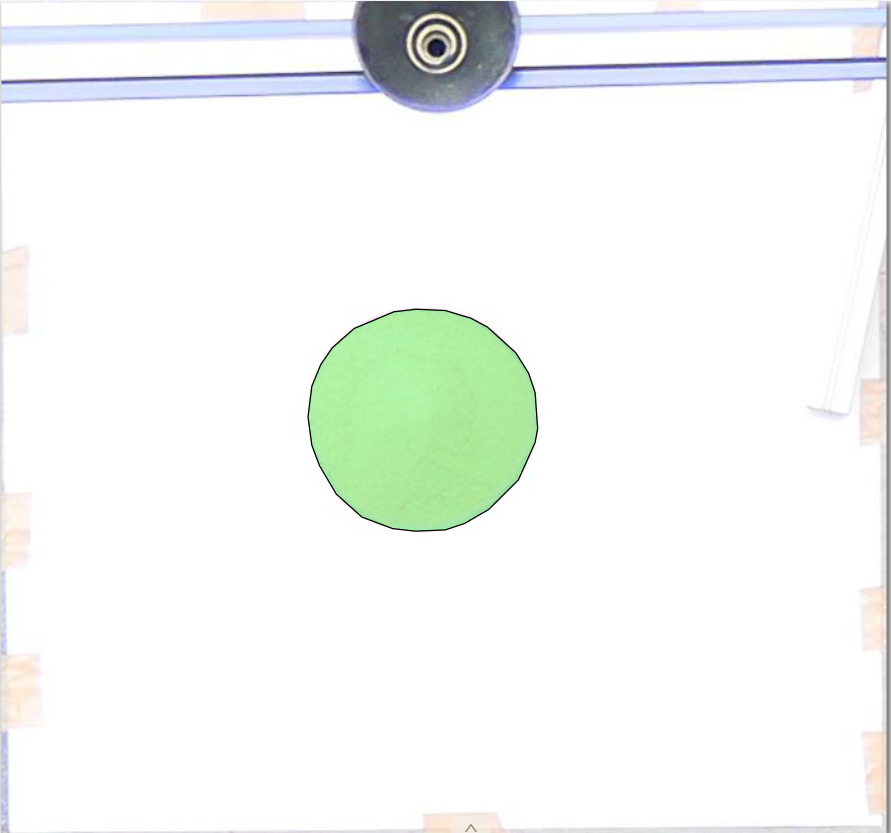}
\includegraphics[width=0.16\columnwidth]{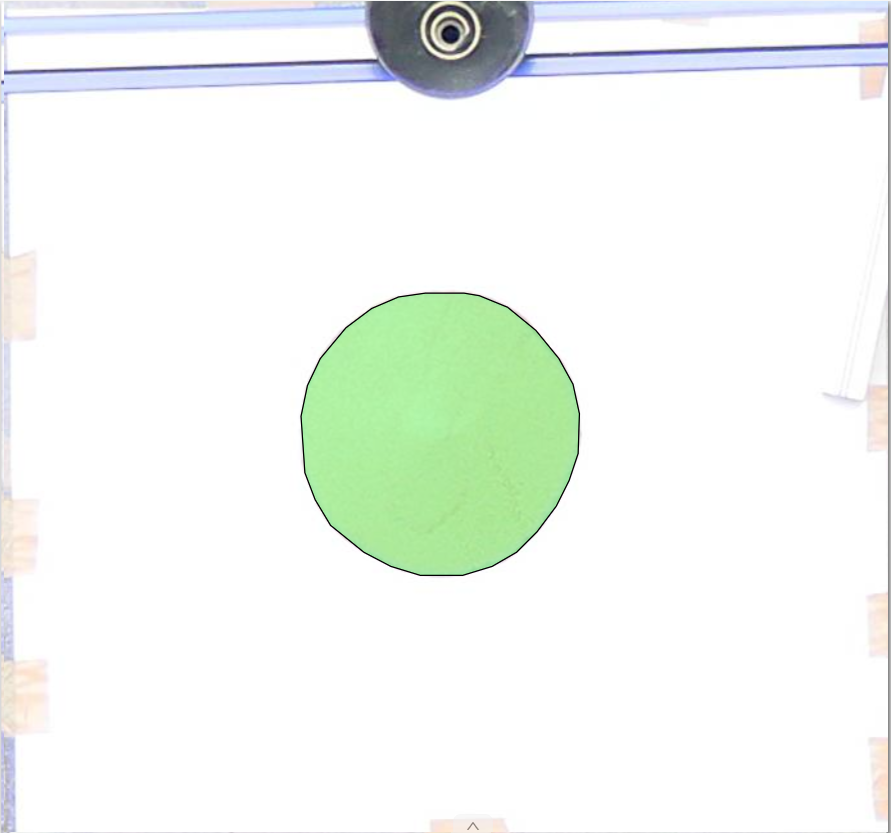}
\includegraphics[width=0.16\columnwidth]{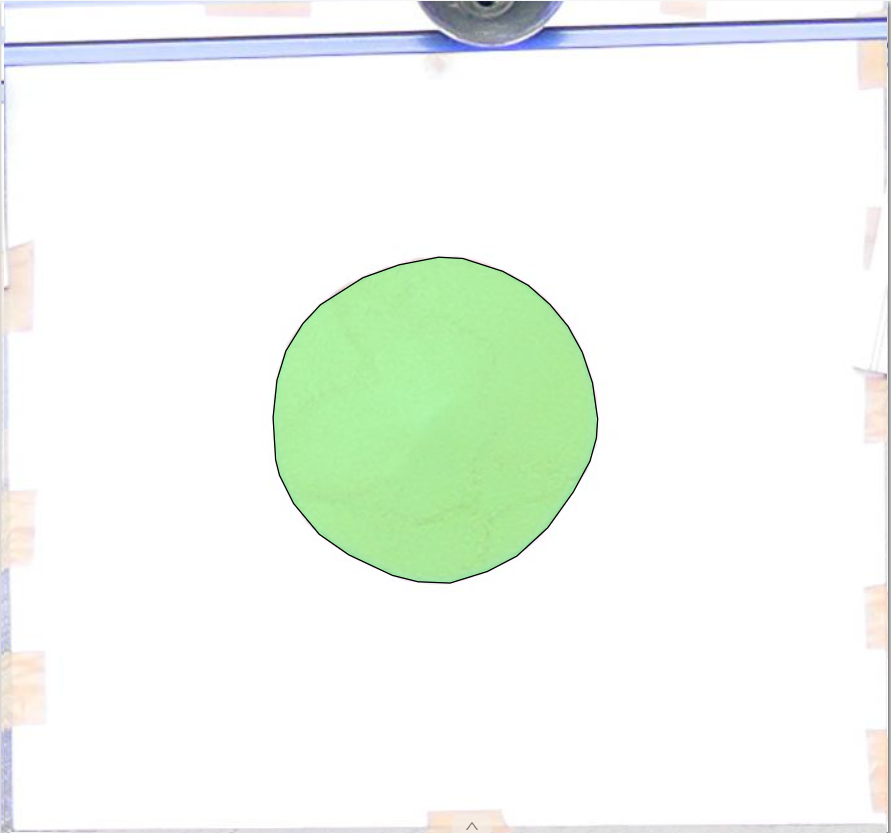}
\includegraphics[width=0.16\columnwidth]{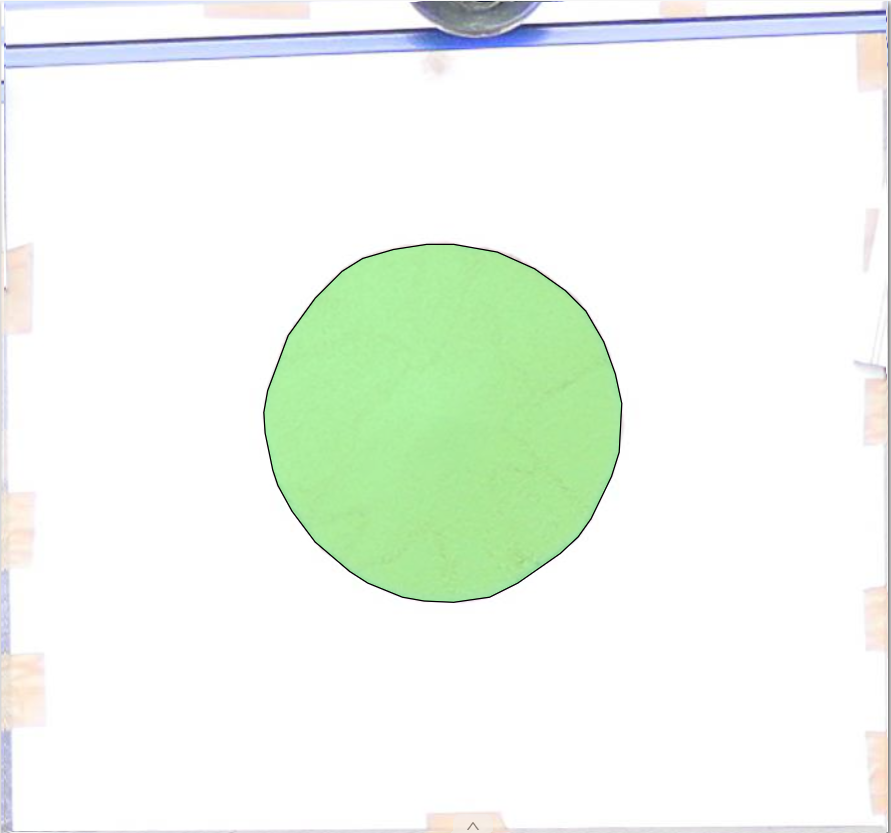}
\includegraphics[width=0.16\columnwidth]{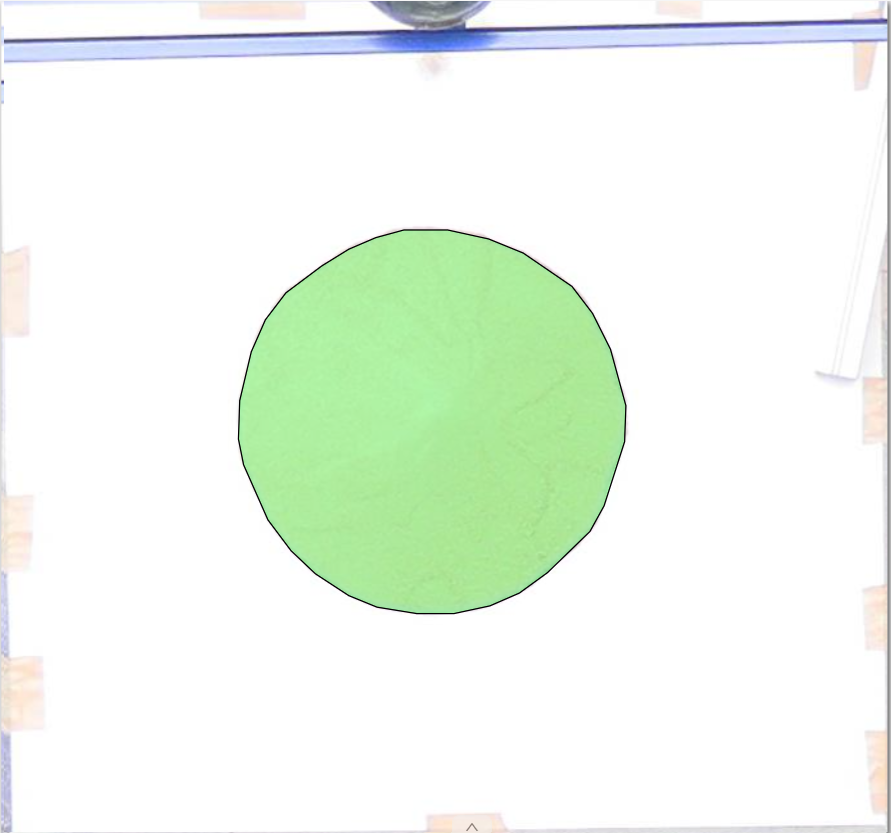}
\includegraphics[width=0.16\columnwidth]{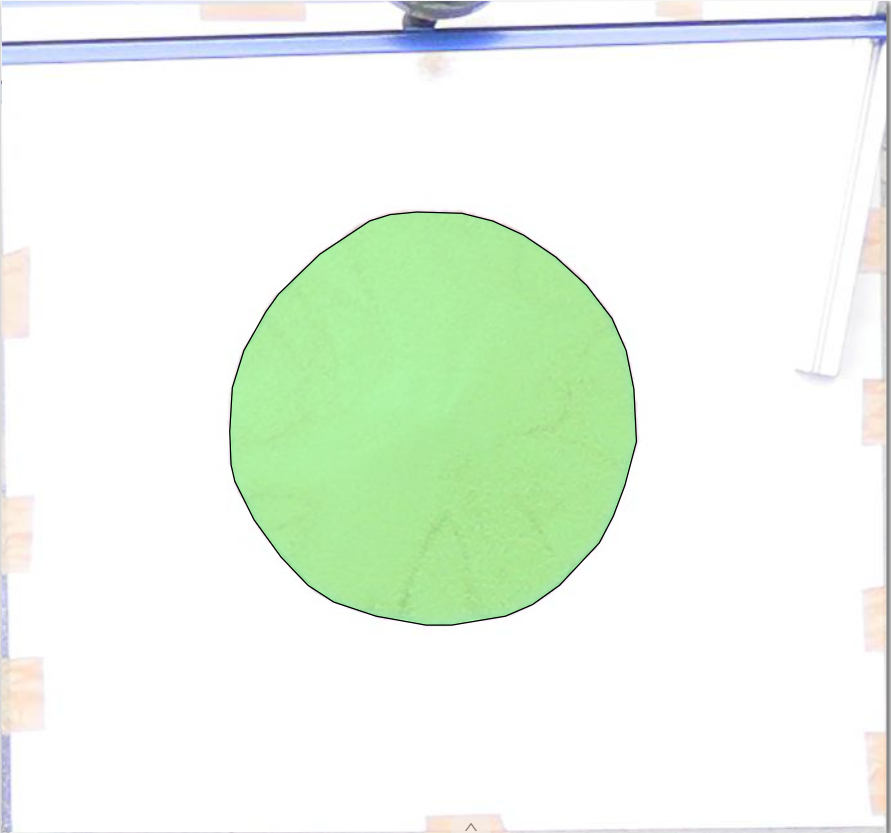}

\includegraphics[width=0.152\textwidth]{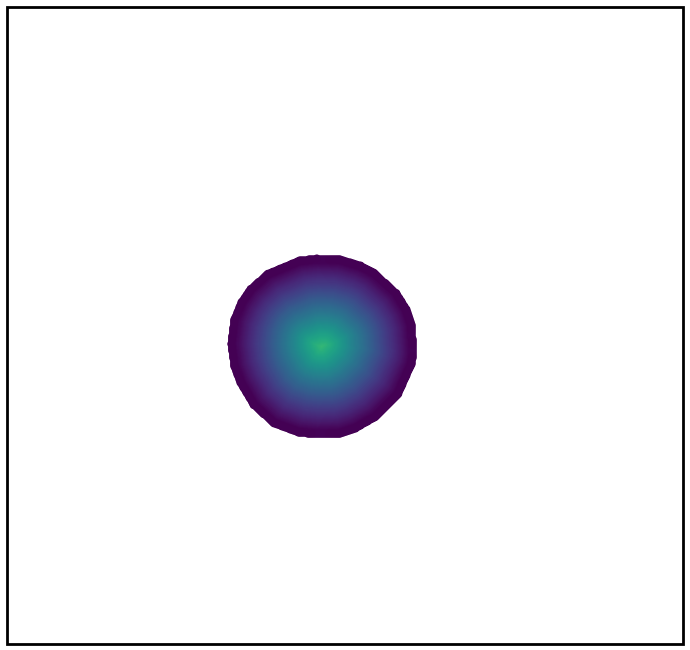}
\includegraphics[width=0.152\textwidth]{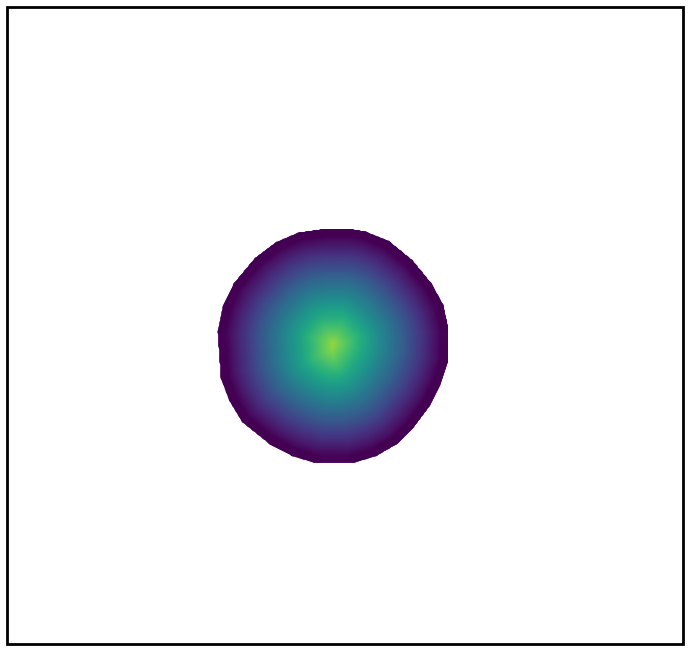}
\includegraphics[width=0.152\textwidth]{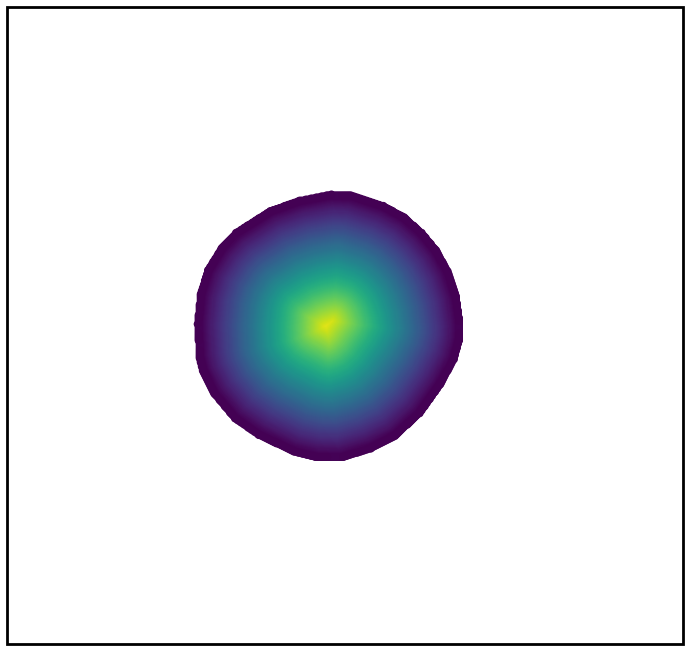}
\includegraphics[width=0.152\textwidth]{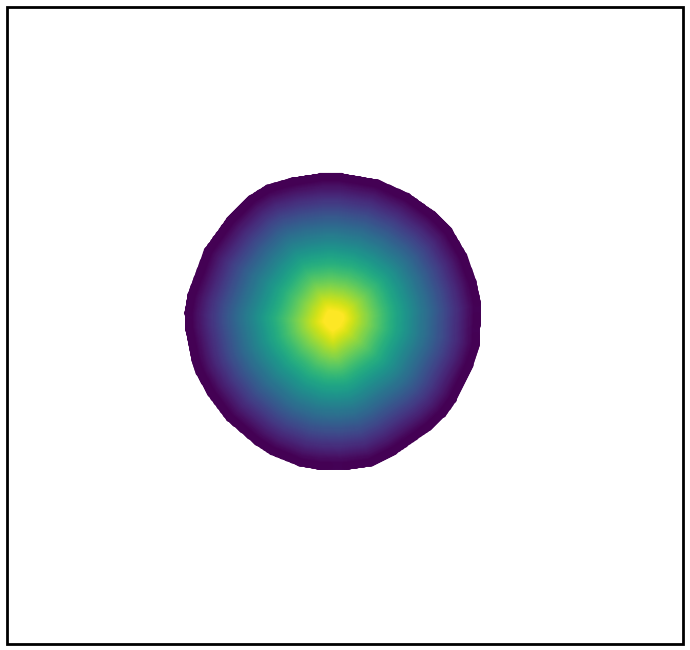}
\includegraphics[width=0.152\textwidth]{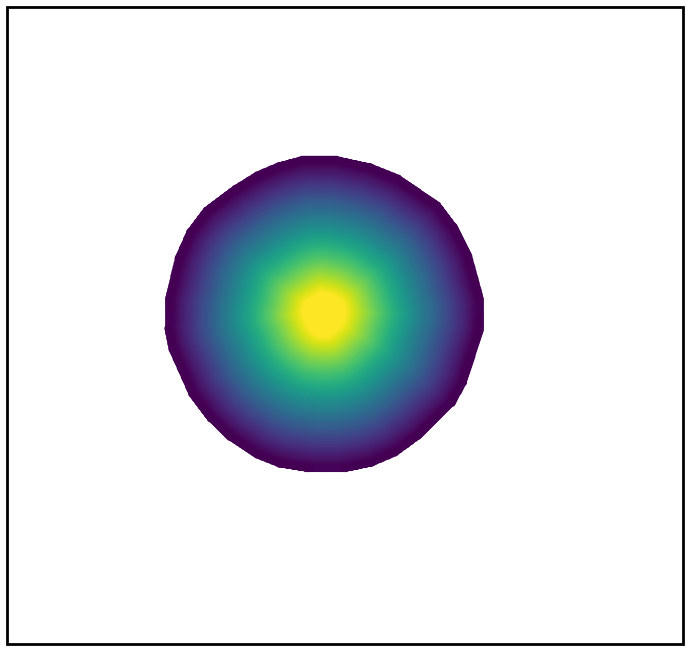}
\includegraphics[width=0.152\textwidth]{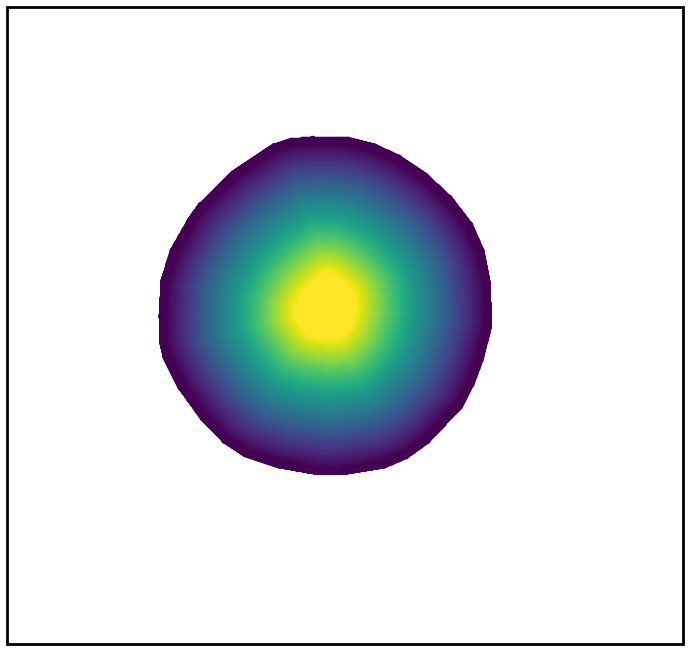}
\includegraphics[width=0.04\textwidth, height= 2.45cm]{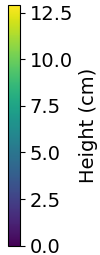}

\caption{Conical full (top), segmentation masks (middle) and  conical full reconstructed (bottom) piles. The first image shows a scale bar (all images have the same scale bar, so  it is not shown in the rest of the images). Each label represents \SI{10}{\centi\meter}. So, the full scale is \SI{90}{\centi\meter}.}
\label{fig:conical_piles_full}
\end{figure*}


\begin{figure*}[!htb]
\centering
\includegraphics[width=0.16\columnwidth]{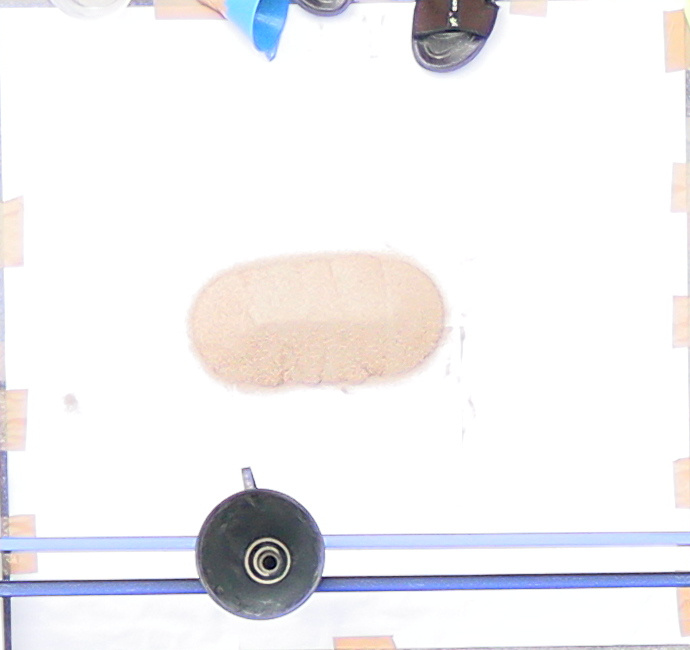}
\includegraphics[width=0.16\columnwidth]{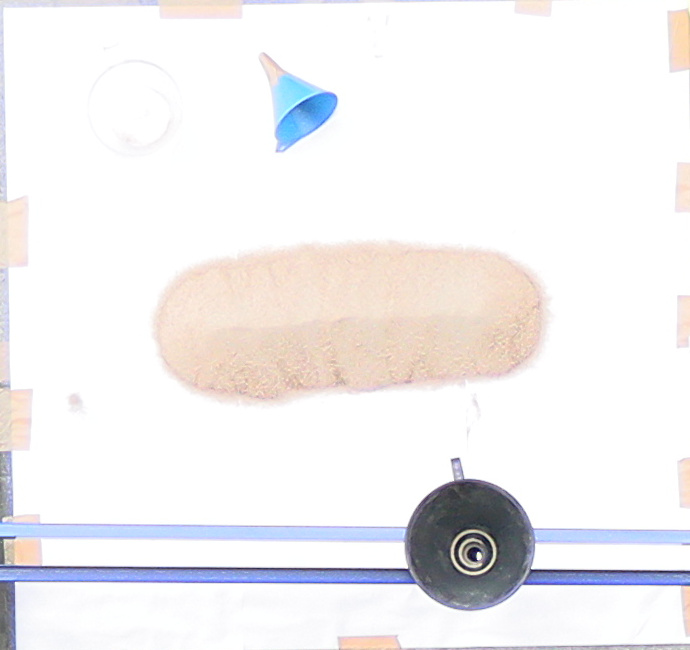}
\includegraphics[width=0.16\columnwidth]{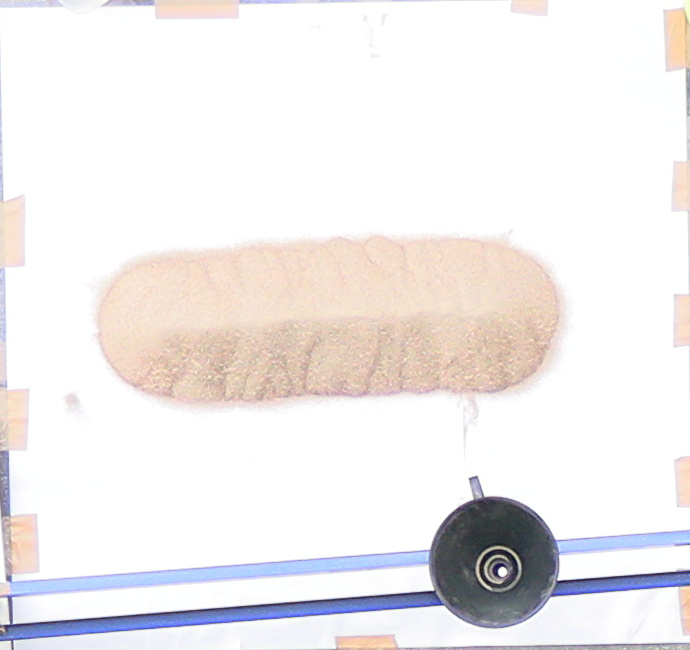}
\includegraphics[width=0.16\columnwidth]{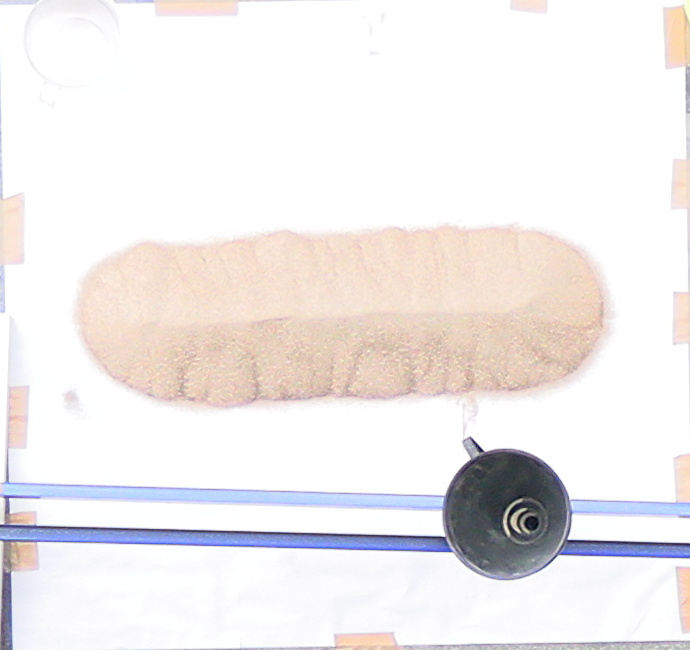}
\includegraphics[width=0.16\columnwidth]{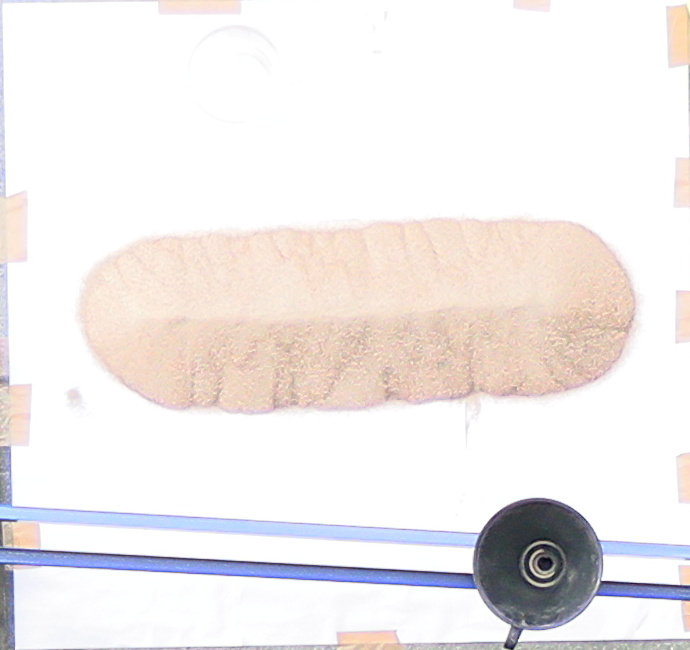}
\includegraphics[width=0.16\columnwidth]{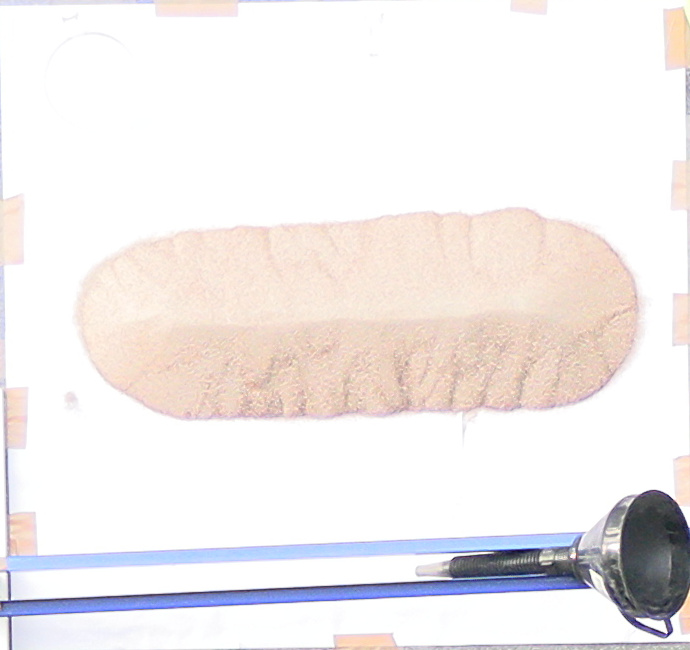}

\includegraphics[width=0.16\columnwidth]{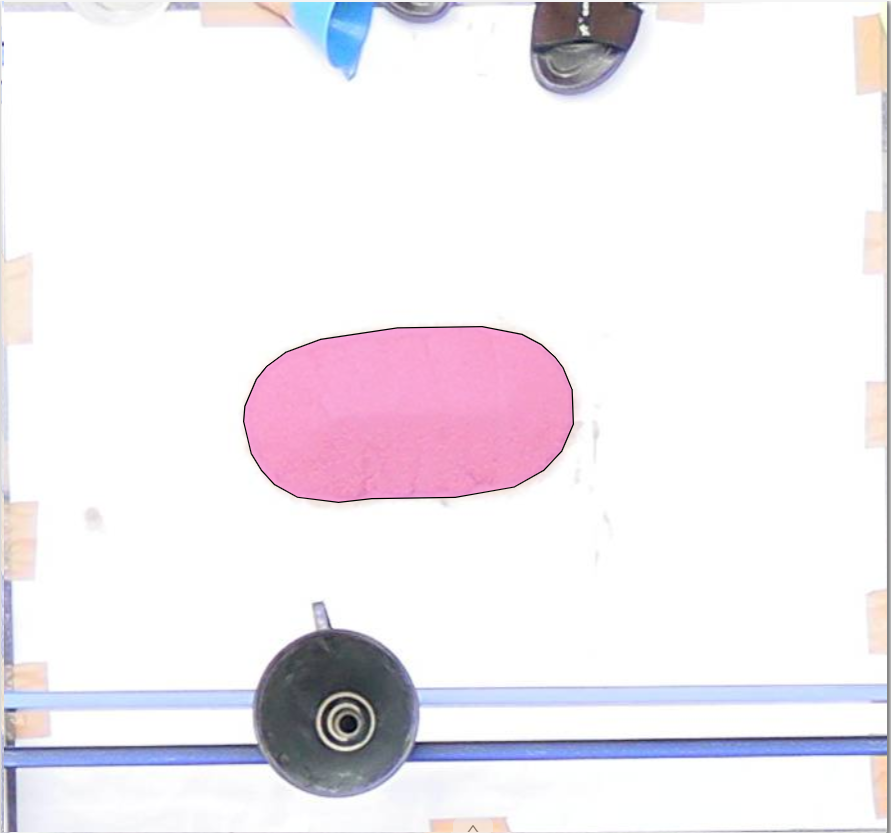}
\includegraphics[width=0.16\columnwidth]{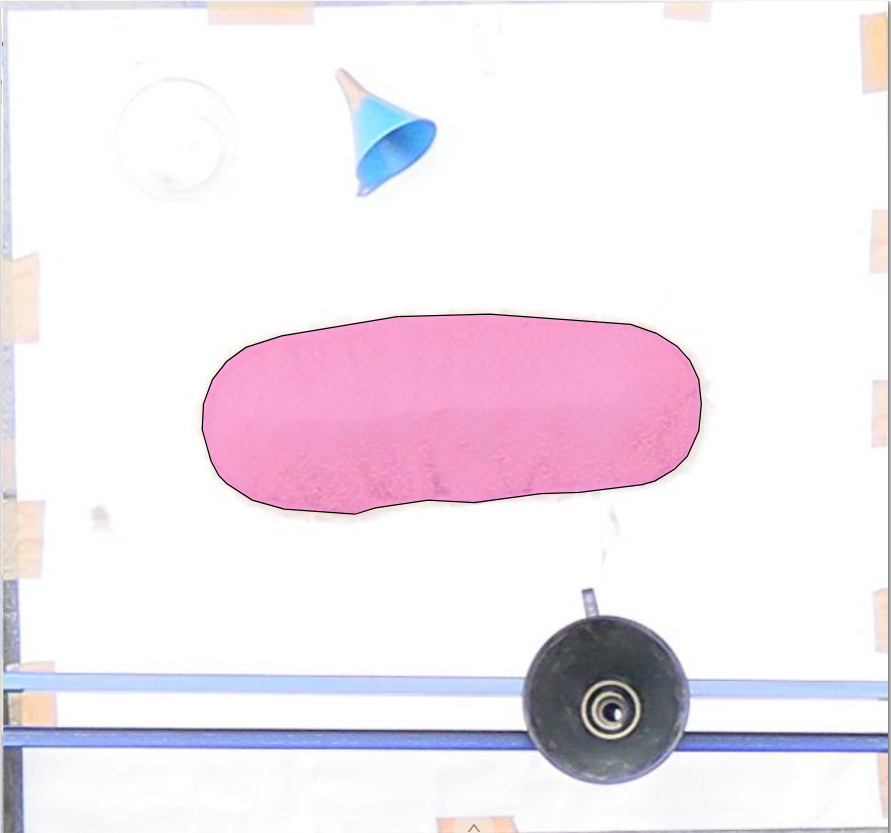}
\includegraphics[width=0.16\columnwidth]{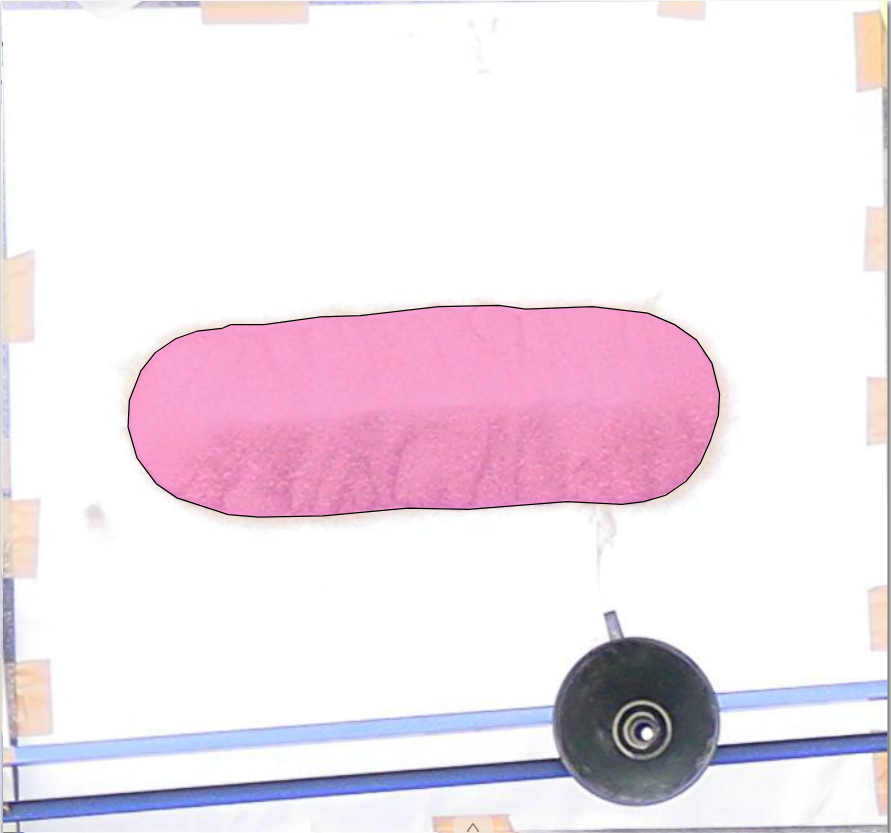}
\includegraphics[width=0.16\columnwidth]{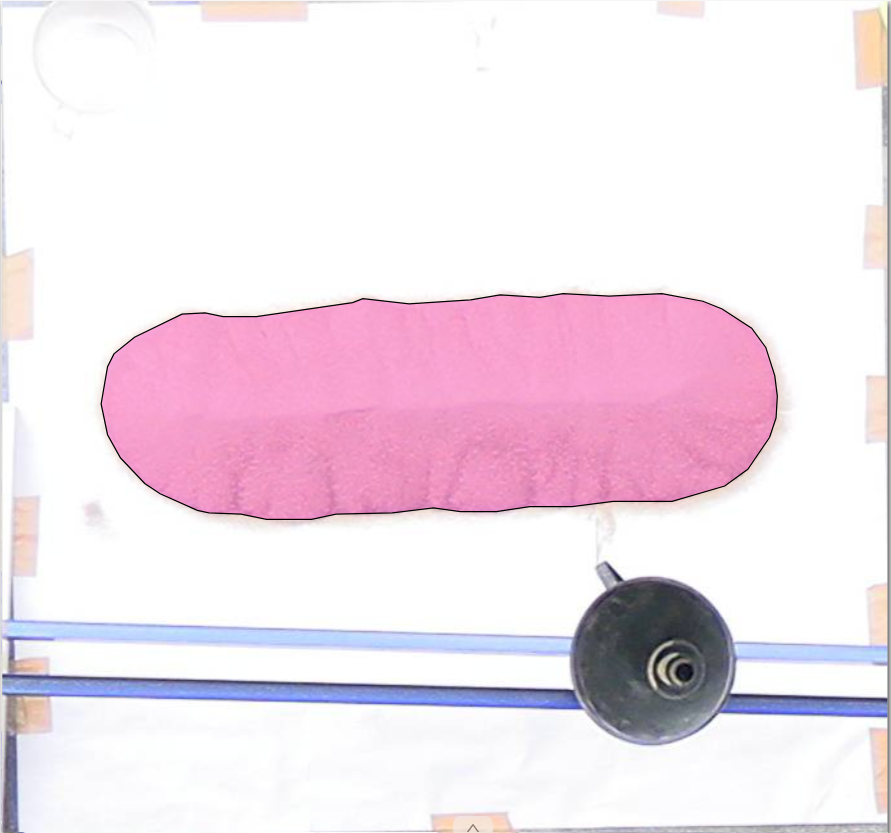}
\includegraphics[width=0.16\columnwidth]{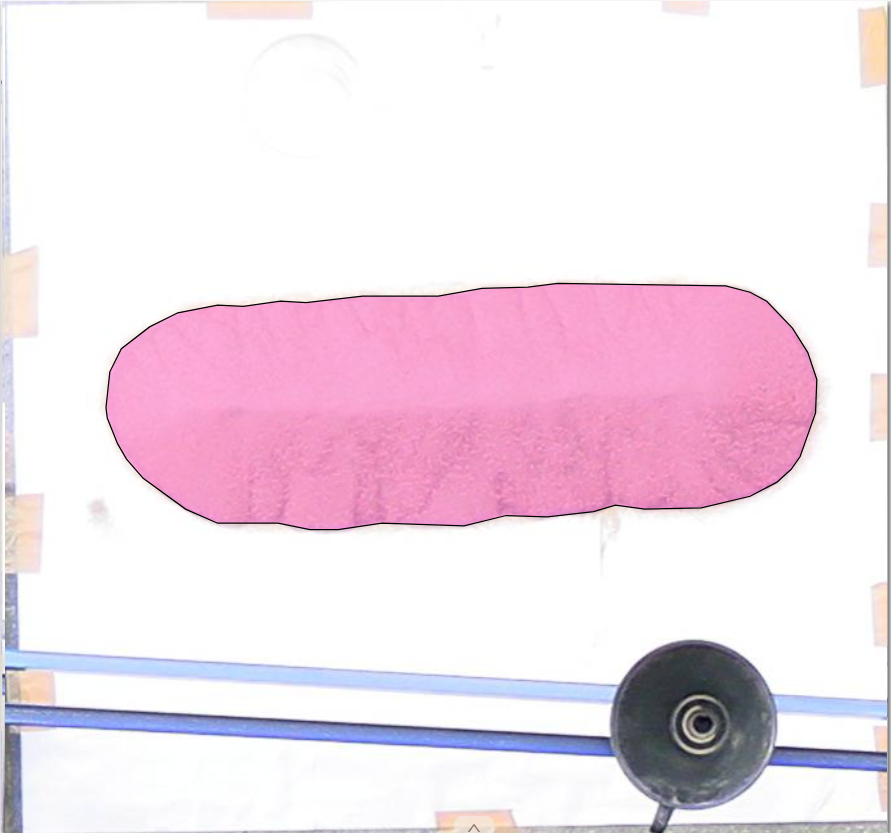}
\includegraphics[width=0.16\columnwidth]{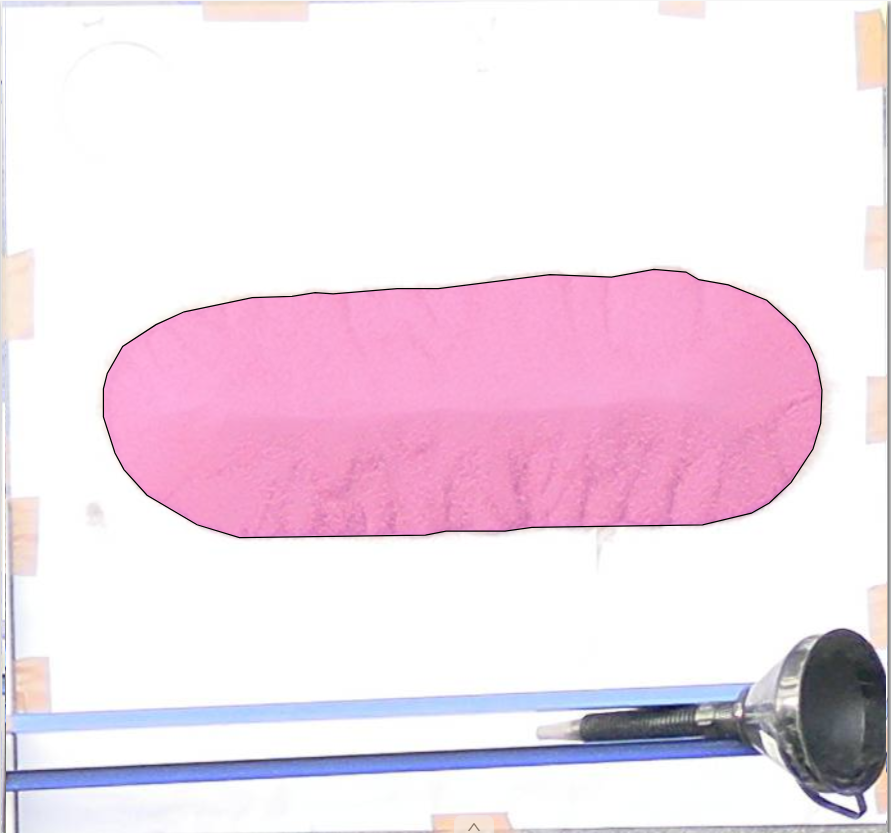}

\includegraphics[width=0.152\textwidth]{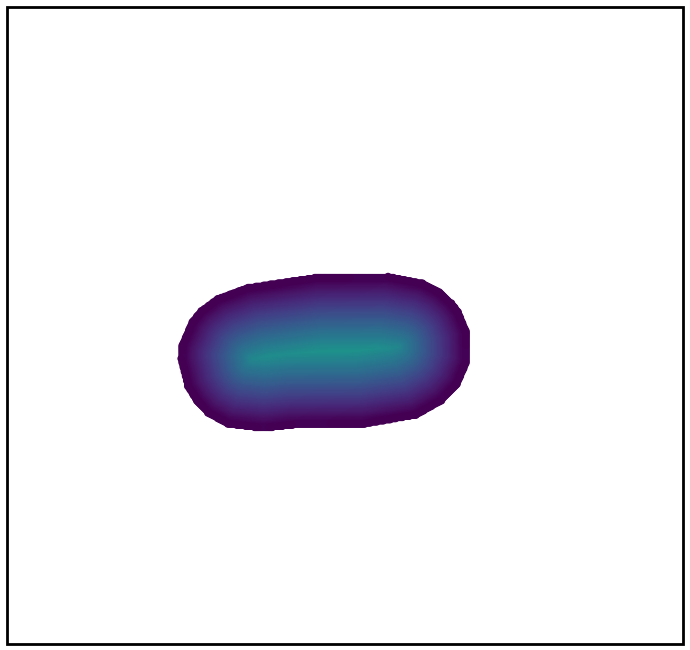}
\includegraphics[width=0.152\textwidth]{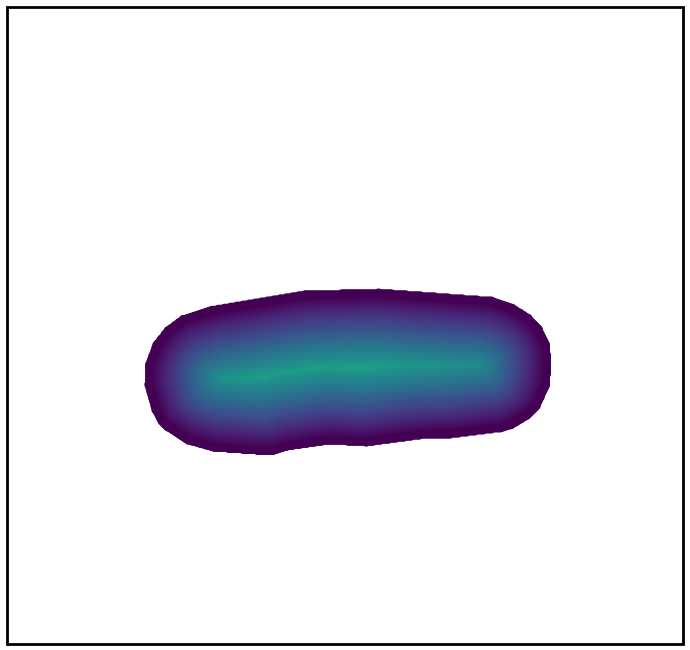}
\includegraphics[width=0.152\textwidth]{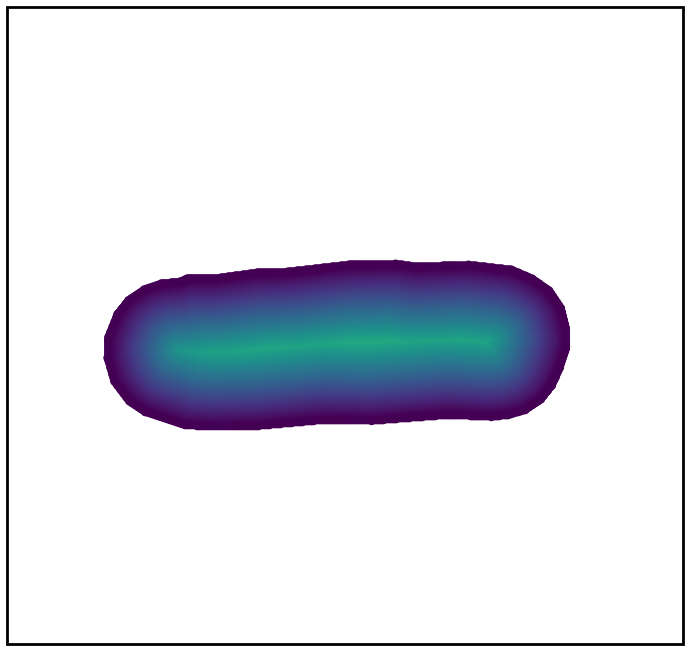}
\includegraphics[width=0.152\textwidth]{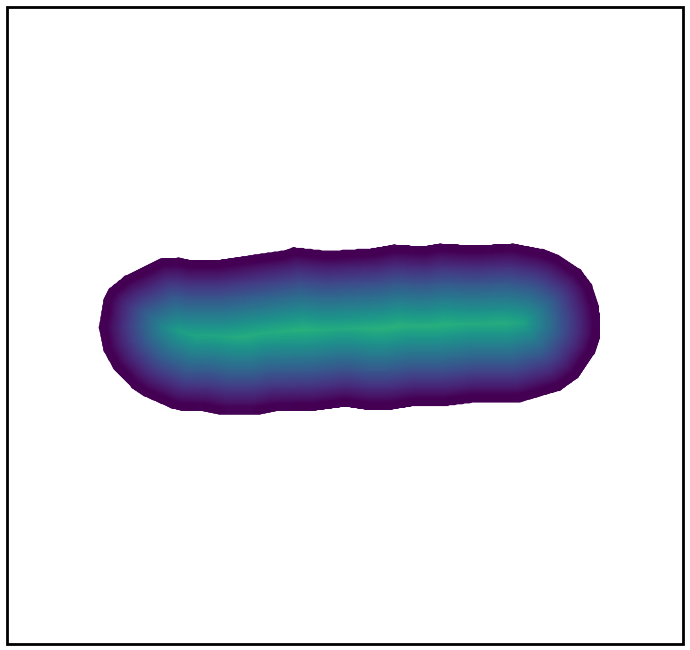}
\includegraphics[width=0.152\textwidth]{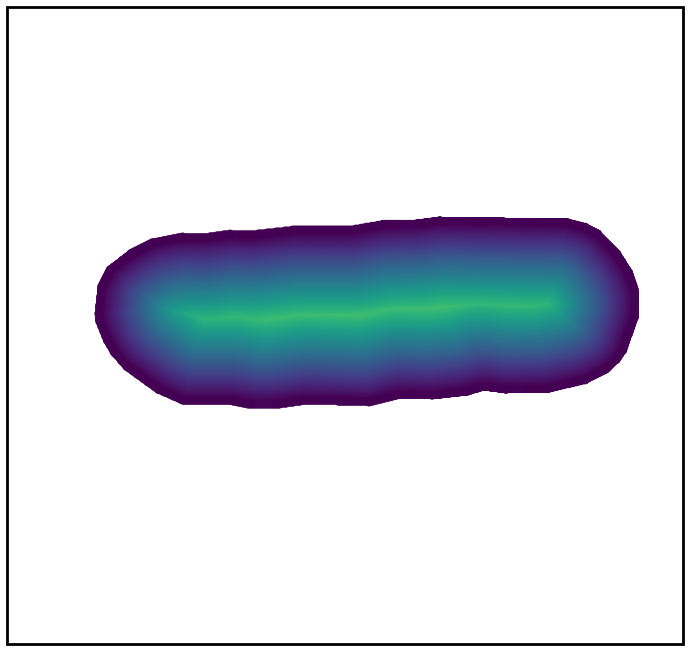}
\includegraphics[width=0.152\textwidth]{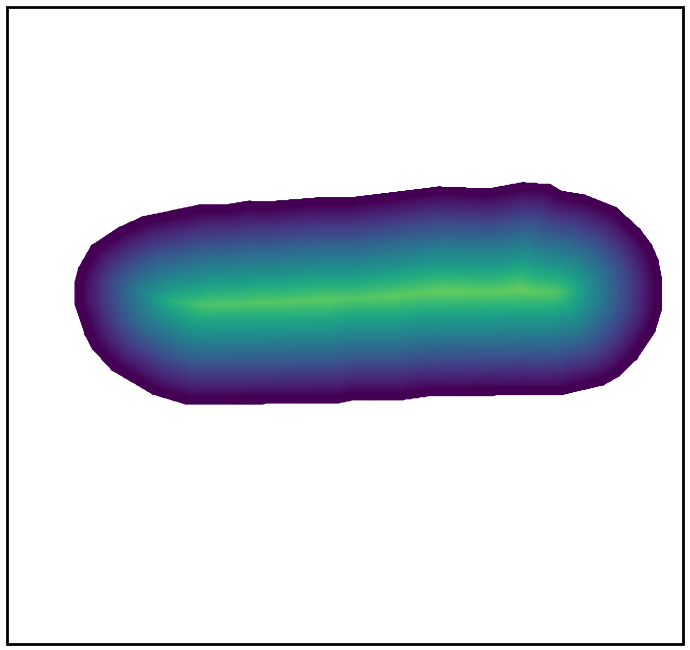}
\includegraphics[width=0.04\textwidth, height = 2.45cm]{colorbar.png}

\caption{Elongated full (top), segmentation masks (middle) and elongated full reconstructed (bottom) piles}
\label{fig:elongated_piles_full}
\end{figure*}

\begin{figure*}[!htb]
\centering
\includegraphics[width=0.16\columnwidth]{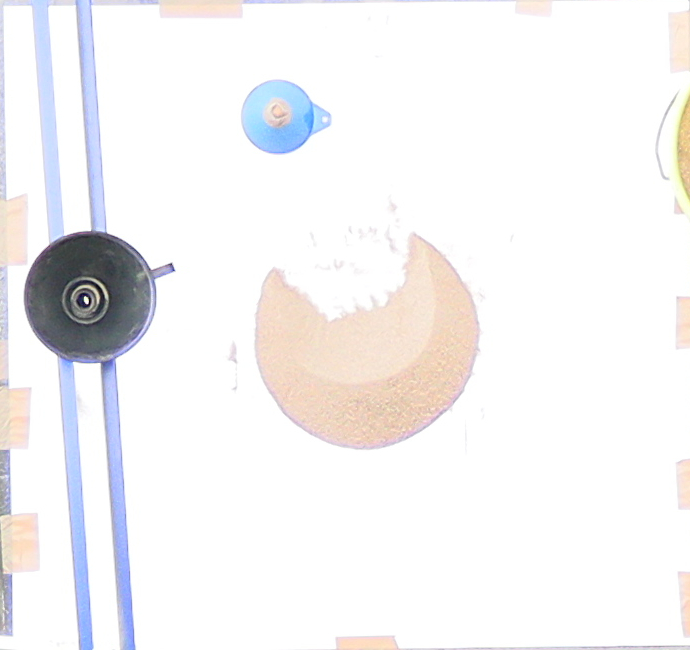}
\includegraphics[width=0.16\columnwidth]{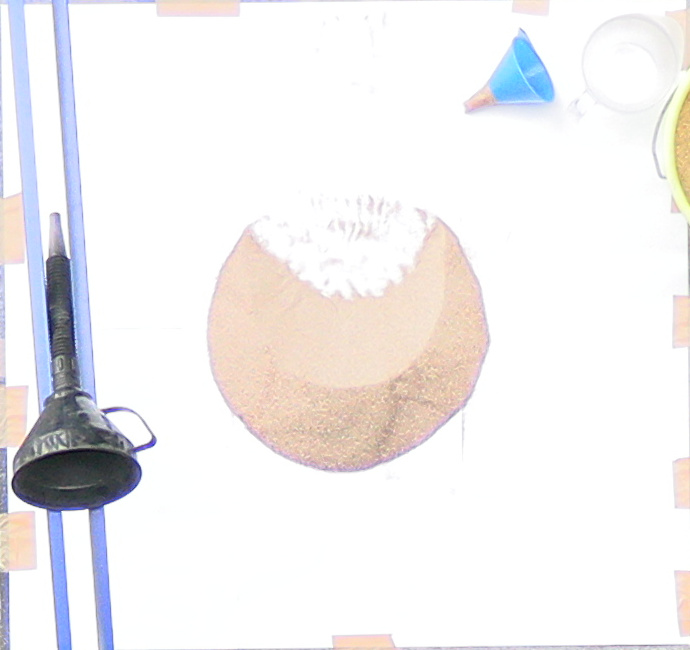}
\includegraphics[width=0.16\columnwidth]{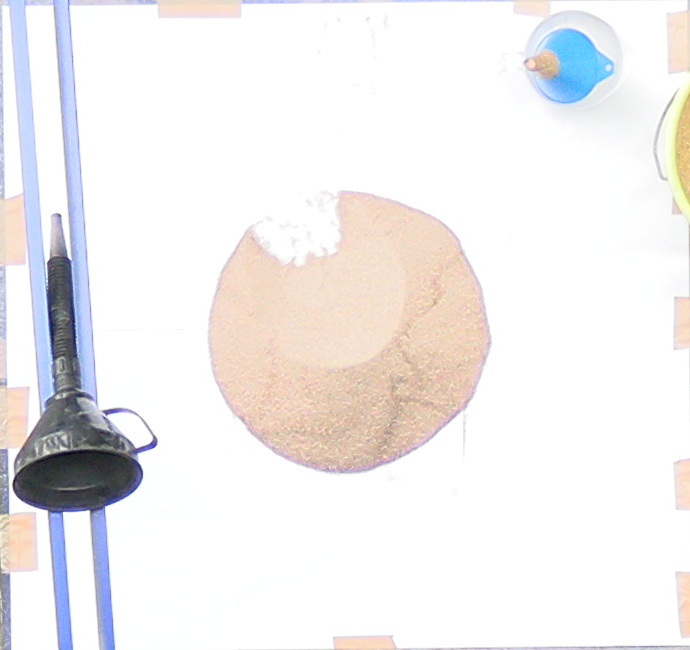}
\includegraphics[width=0.16\columnwidth]{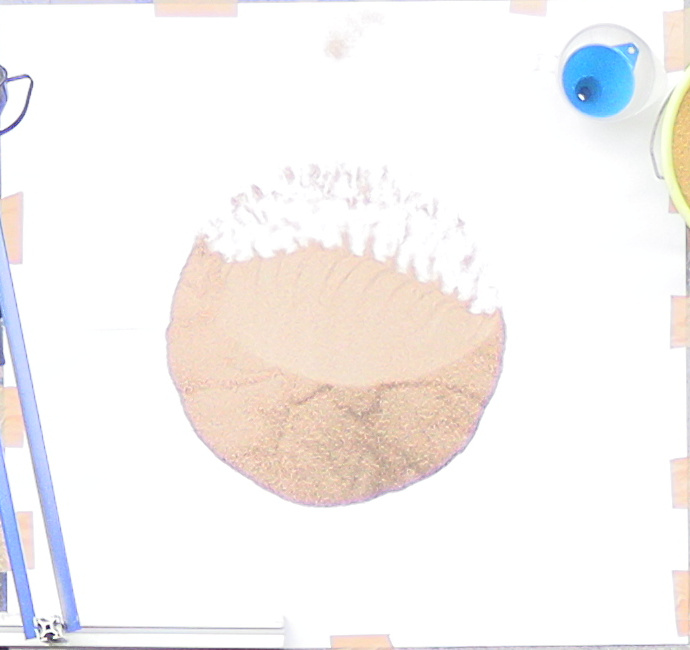}
\includegraphics[width=0.16\columnwidth]{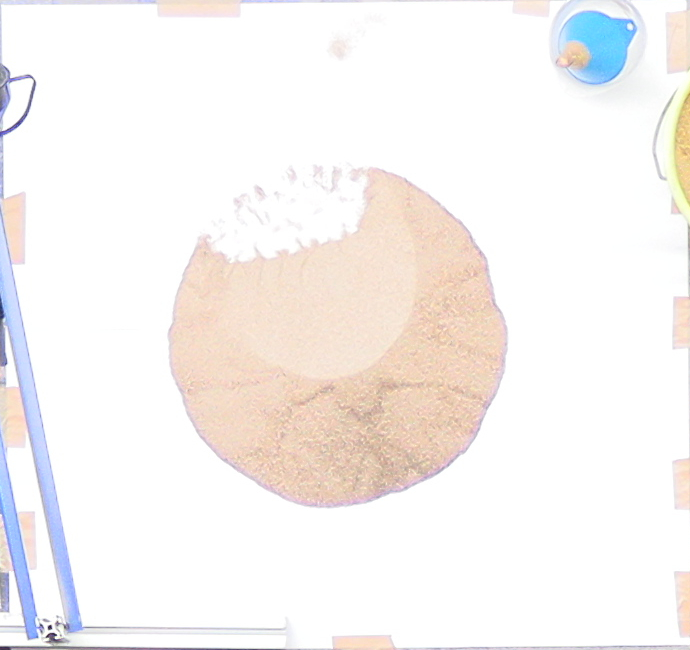}
\includegraphics[width=0.16\columnwidth]{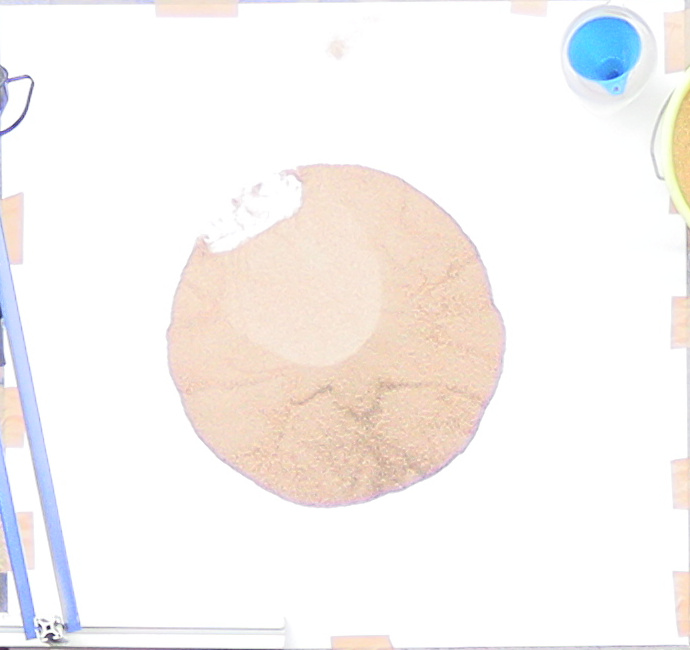}

\includegraphics[width=0.16\columnwidth]{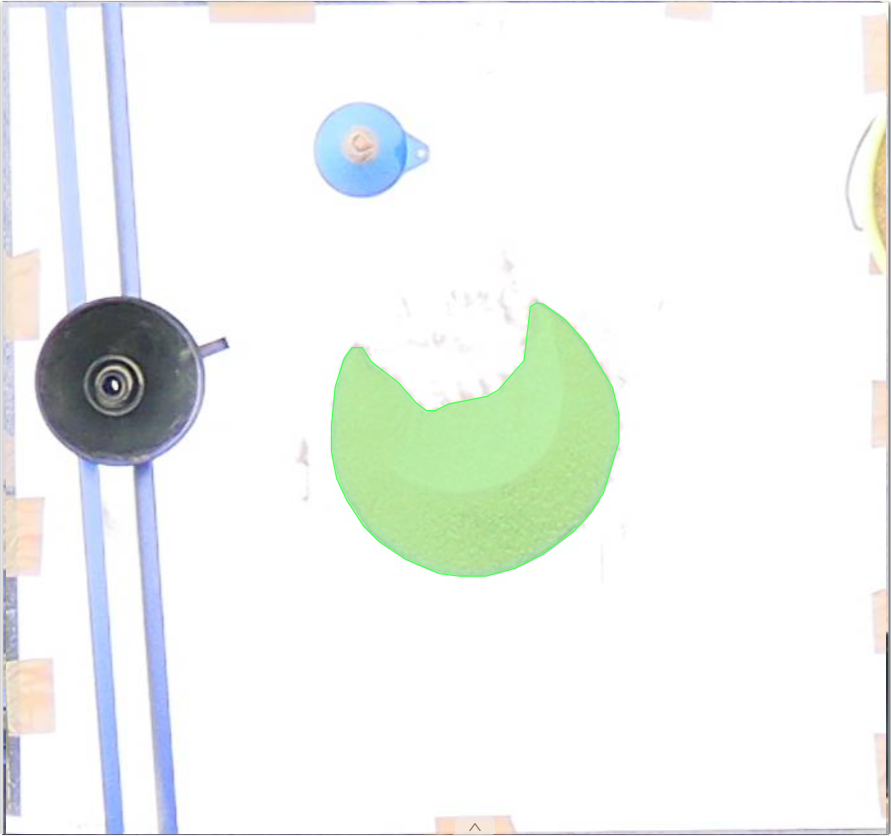}
\includegraphics[width=0.16\columnwidth]{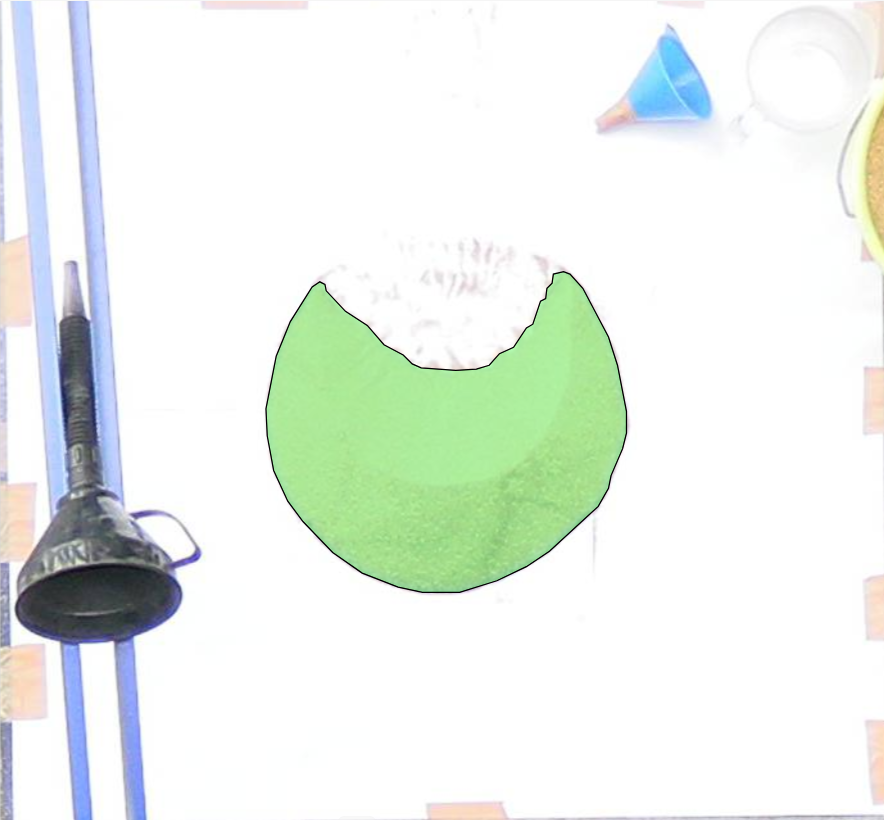}
\includegraphics[width=0.16\columnwidth]{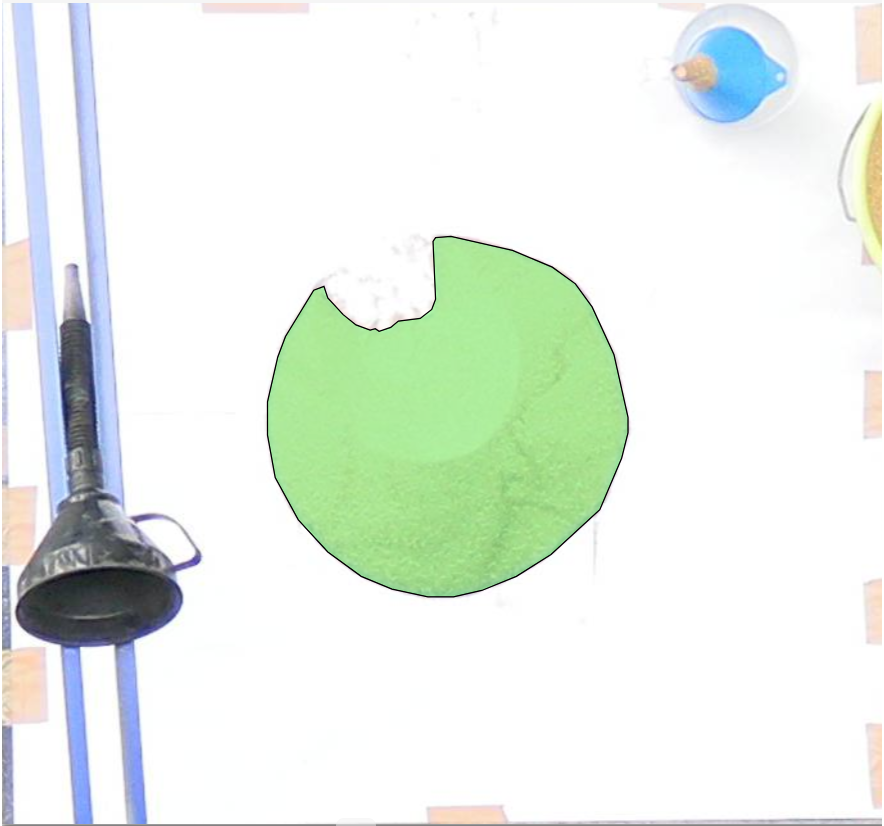}
\includegraphics[width=0.16\columnwidth]{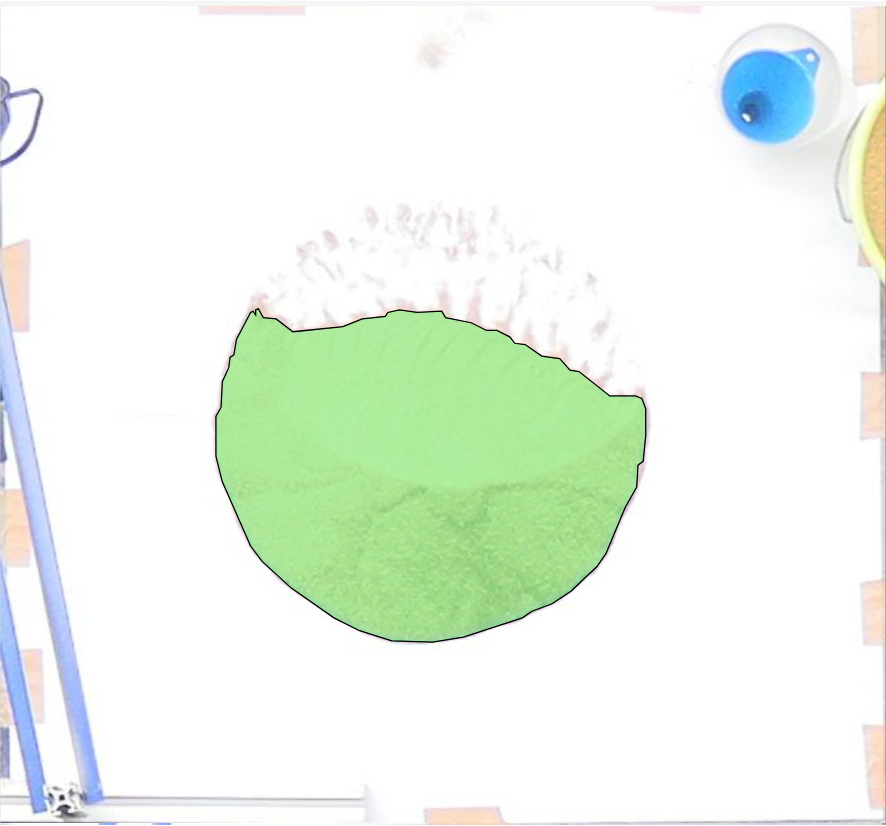}
\includegraphics[width=0.16\columnwidth]{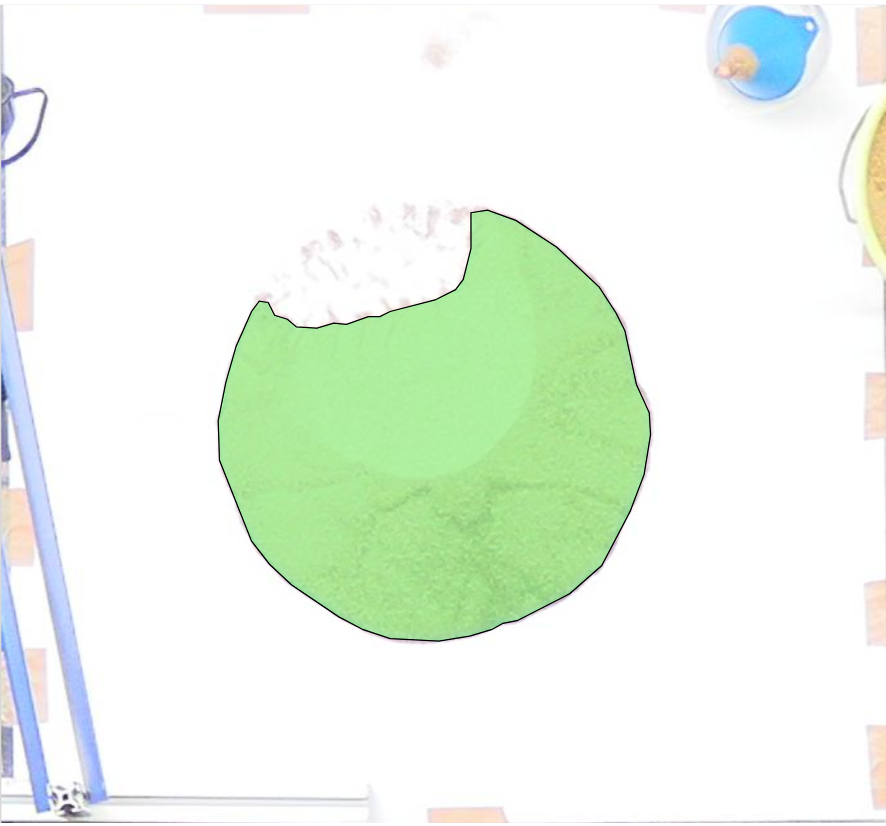}
\includegraphics[width=0.16\columnwidth]{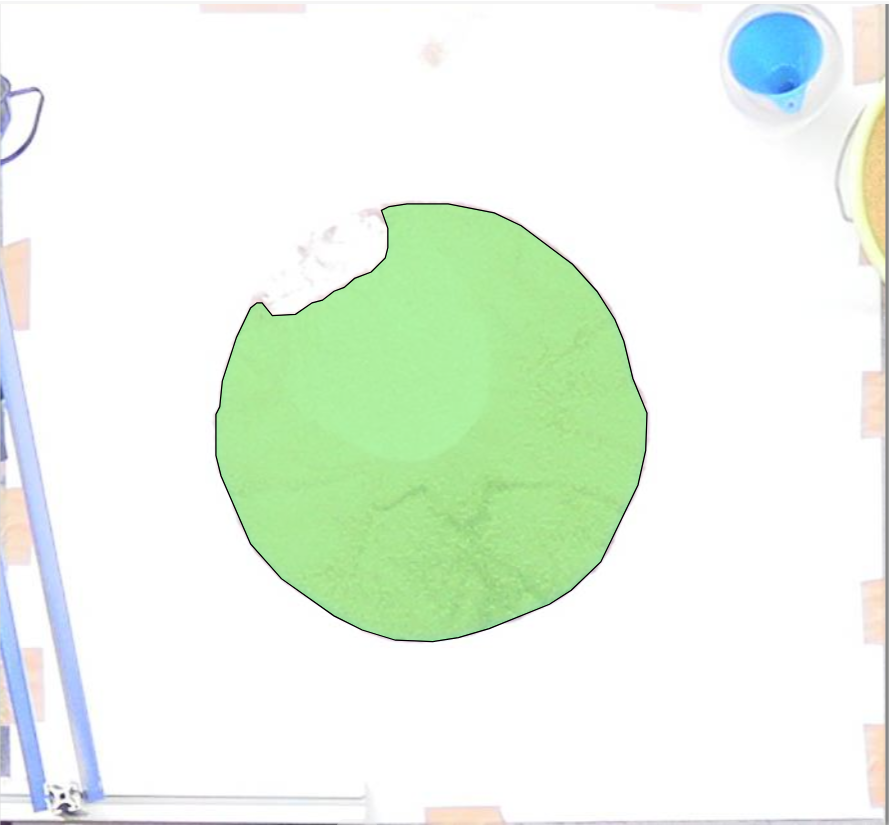}

\includegraphics[width=0.152\textwidth]{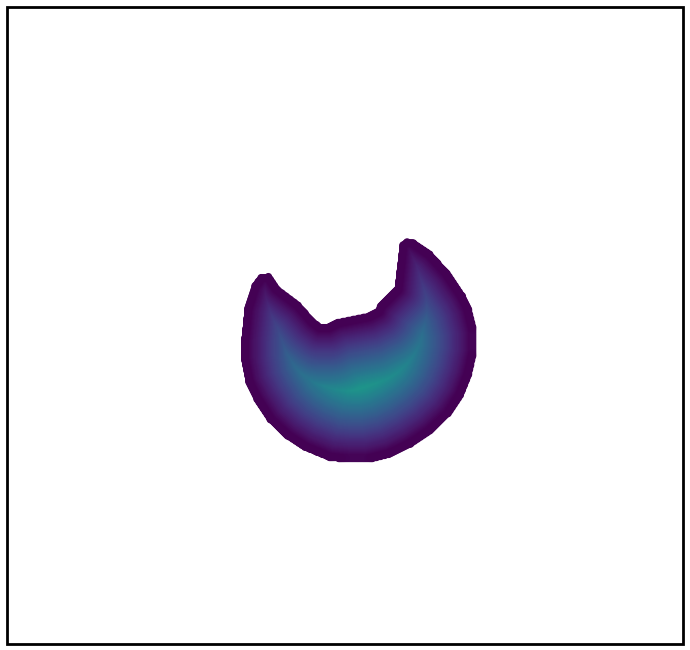}
\includegraphics[width=0.152\textwidth]{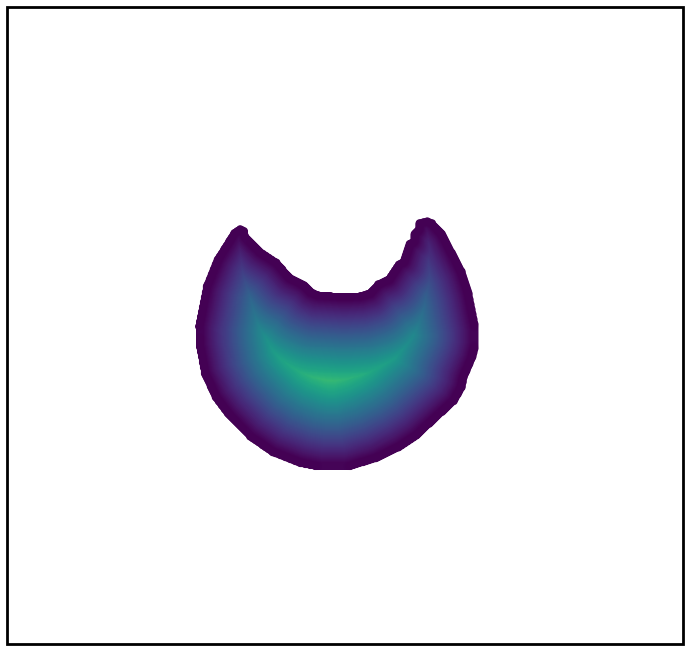}
\includegraphics[width=0.152\textwidth]{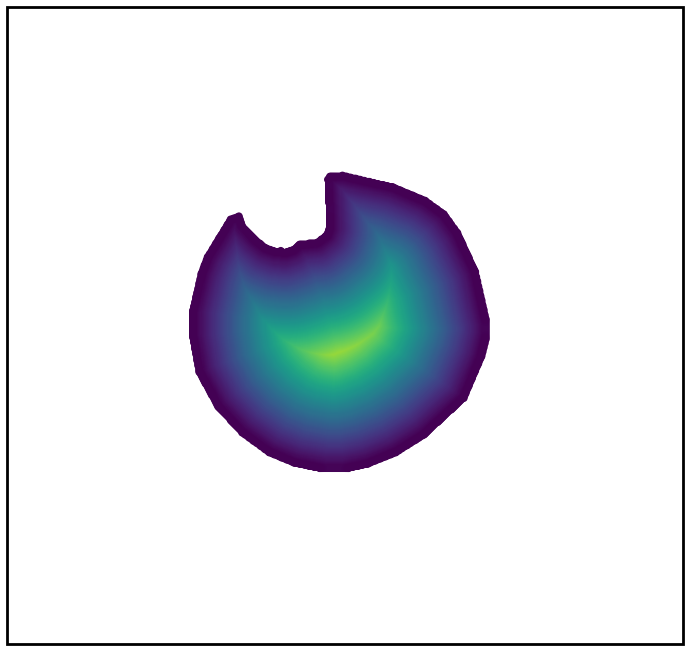}
\includegraphics[width=0.152\textwidth]{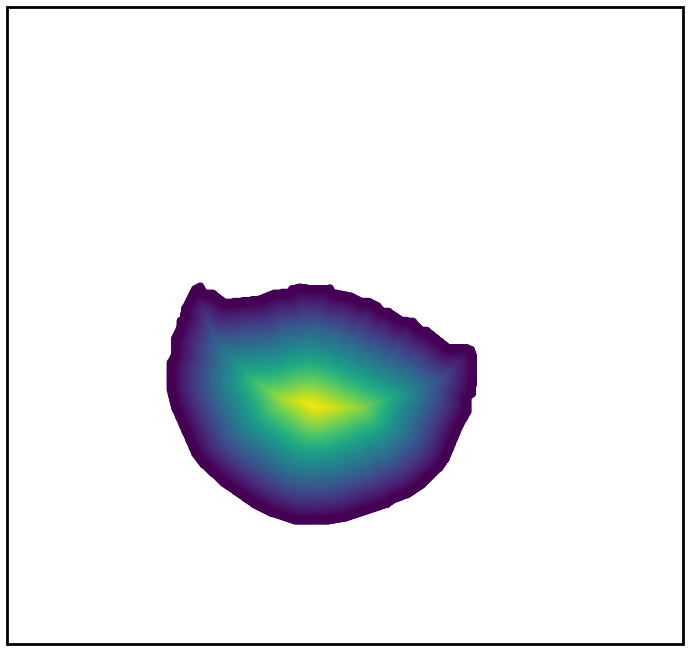}
\includegraphics[width=0.152\textwidth]{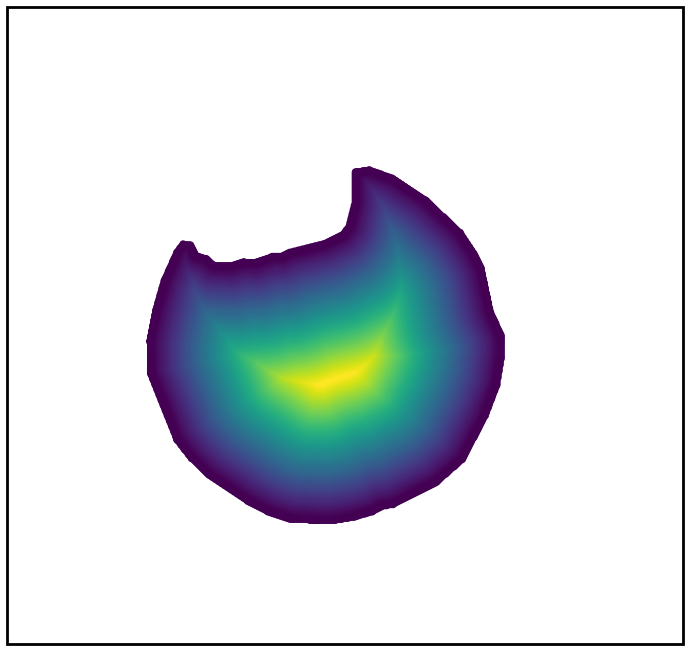}
\includegraphics[width=0.152\textwidth]{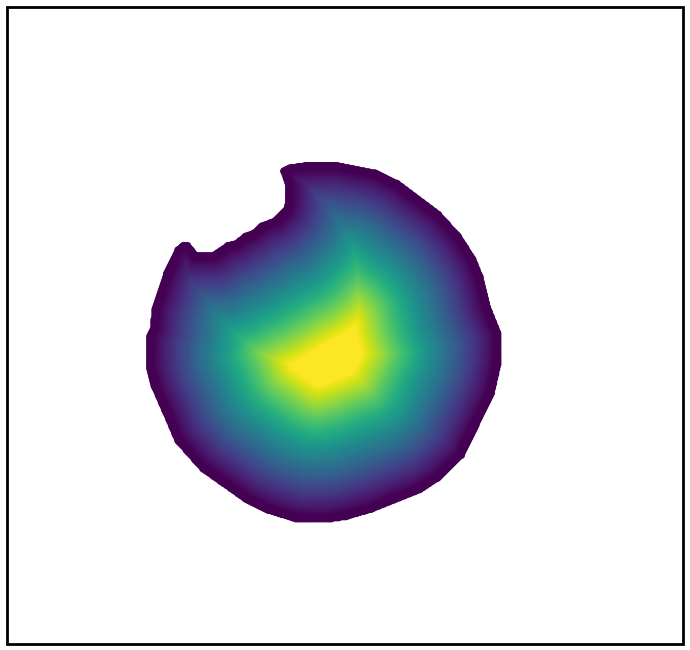}
\includegraphics[width=0.04\textwidth, height = 2.45cm]{colorbar.png}

\caption{Conical reclaimed (top), segmentation masks (middle) and conical reclaimed reconstructed (bottom) piles}
\label{fig:conical_piles_reclaimed}

\end{figure*}

\begin{figure*}[!htb]
\centering
\includegraphics[width=0.16\columnwidth]{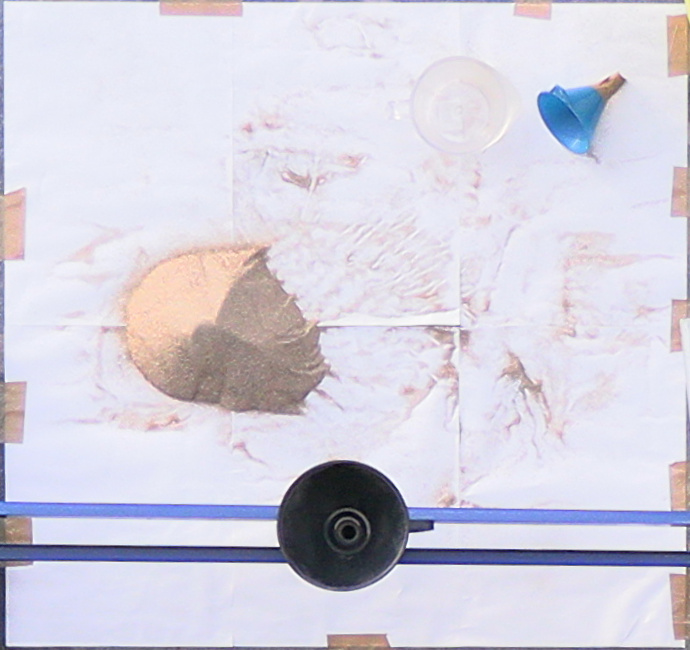}
\includegraphics[width=0.16\columnwidth]{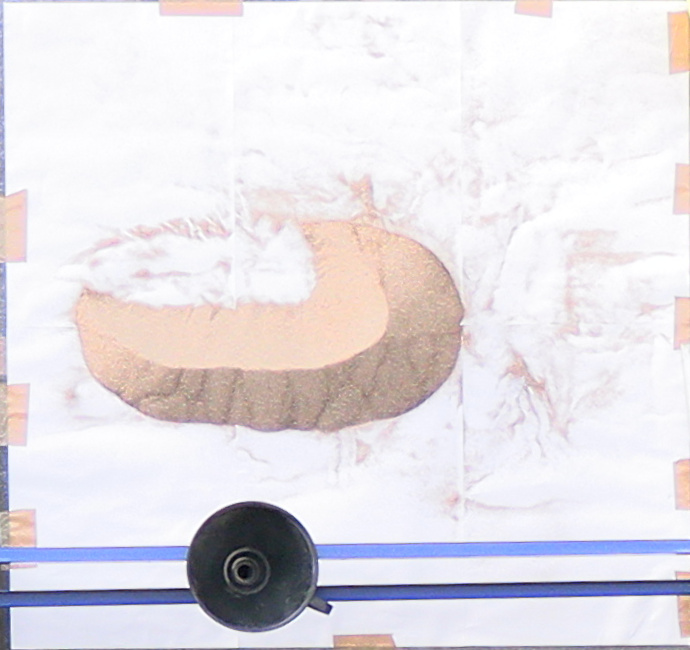}
\includegraphics[width=0.16\columnwidth]{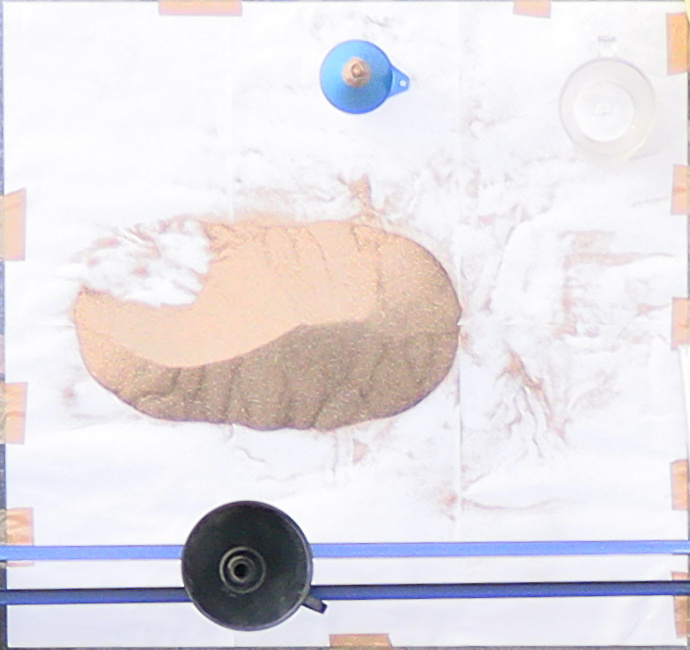}
\includegraphics[width=0.16\columnwidth]{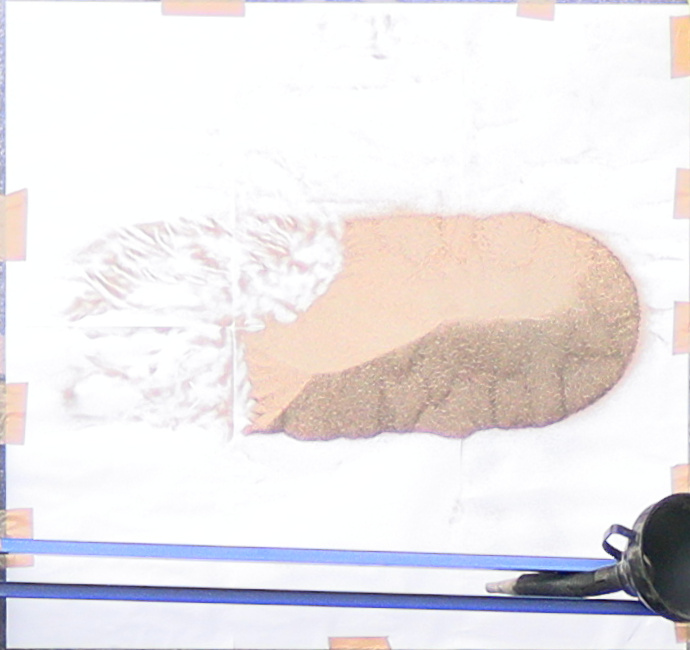}
\includegraphics[width=0.16\columnwidth]{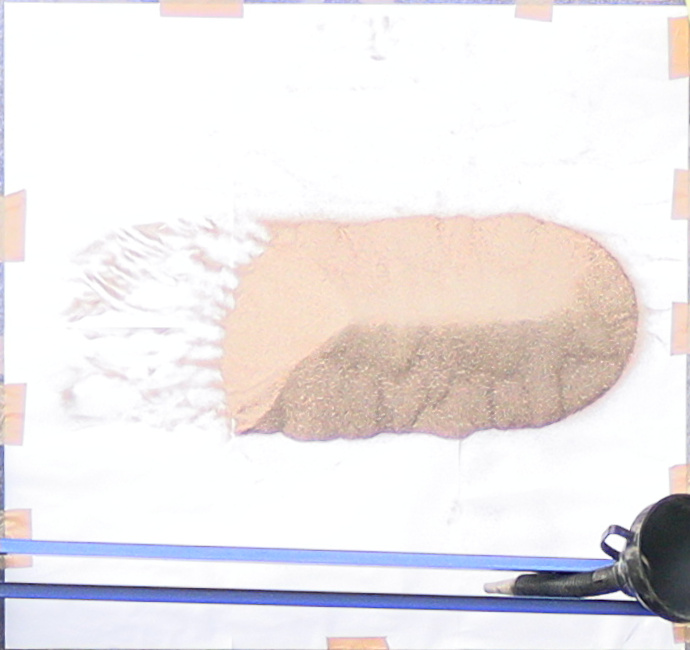}
\includegraphics[width=0.16\columnwidth]{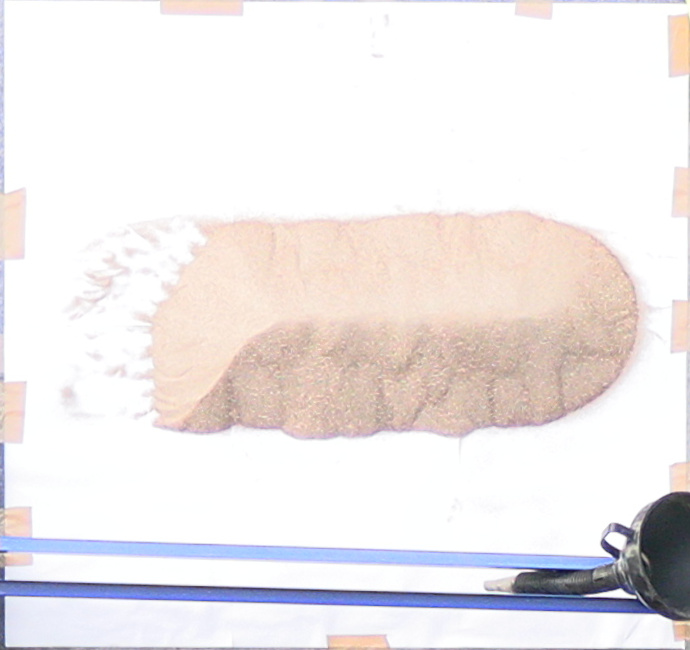}

\includegraphics[width=0.16\columnwidth]{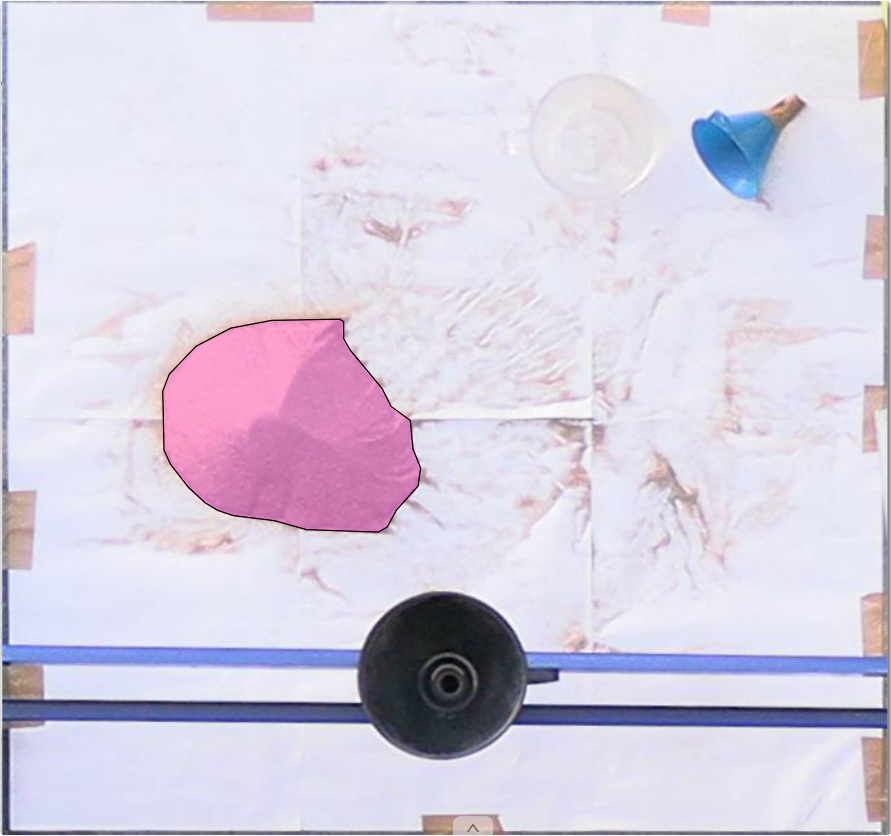}
\includegraphics[width=0.16\columnwidth]{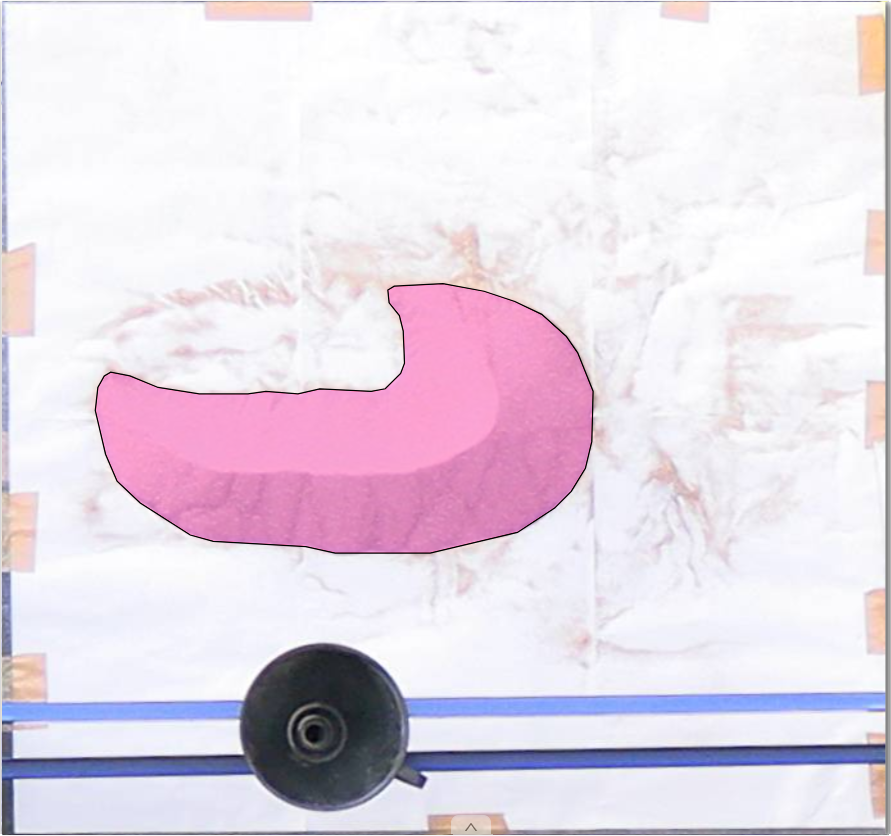}
\includegraphics[width=0.16\columnwidth]{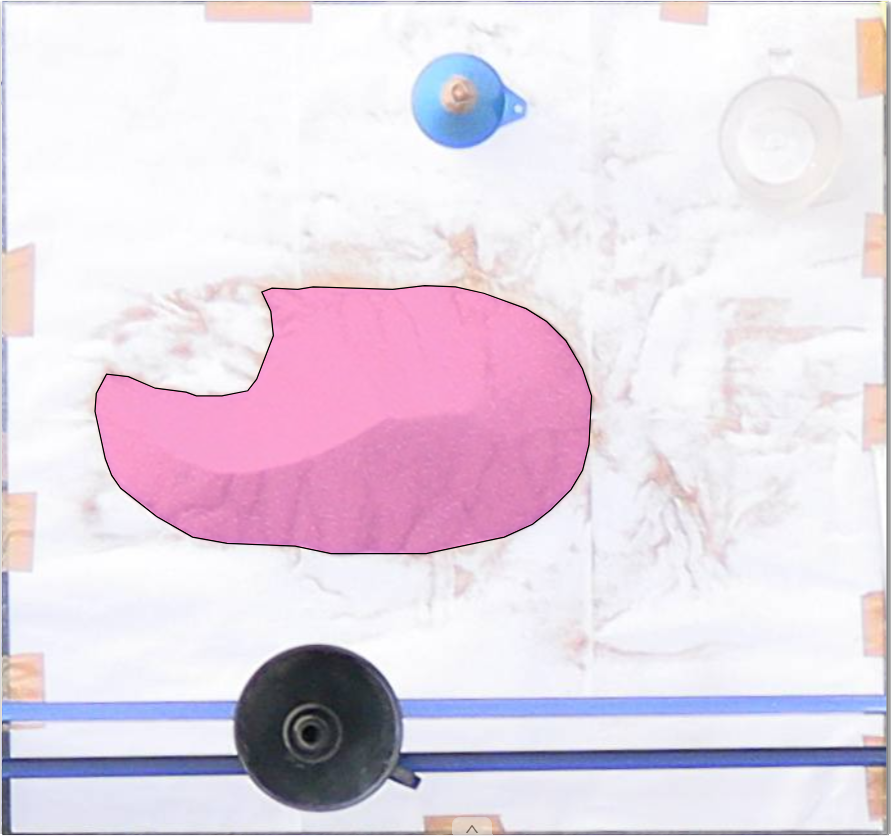}
\includegraphics[width=0.16\columnwidth]{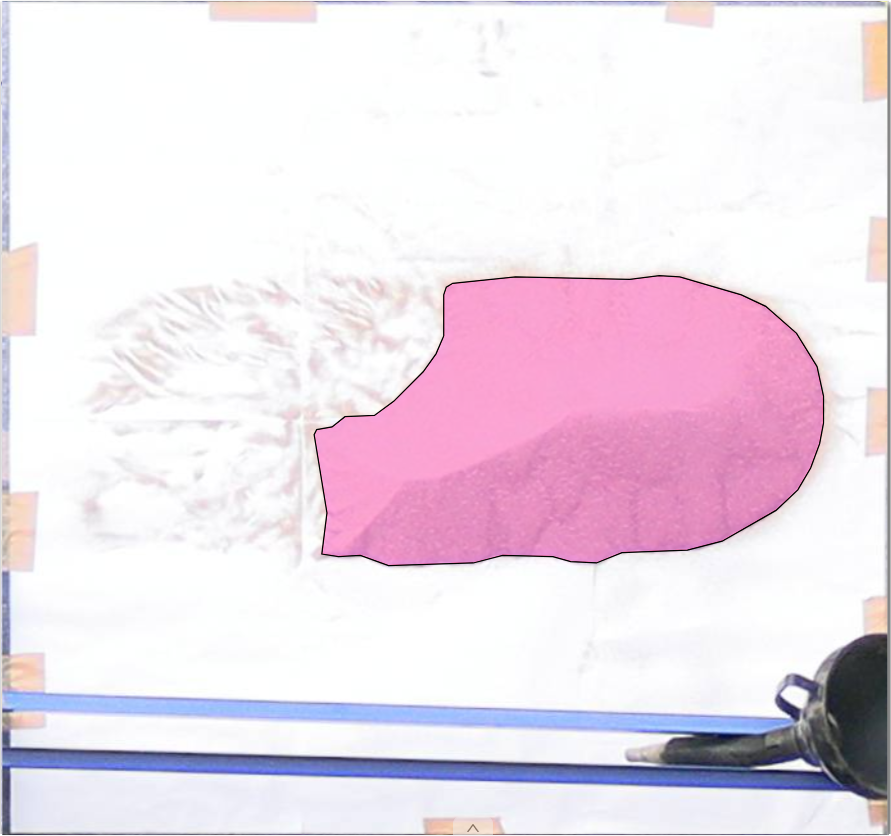}
\includegraphics[width=0.16\columnwidth]{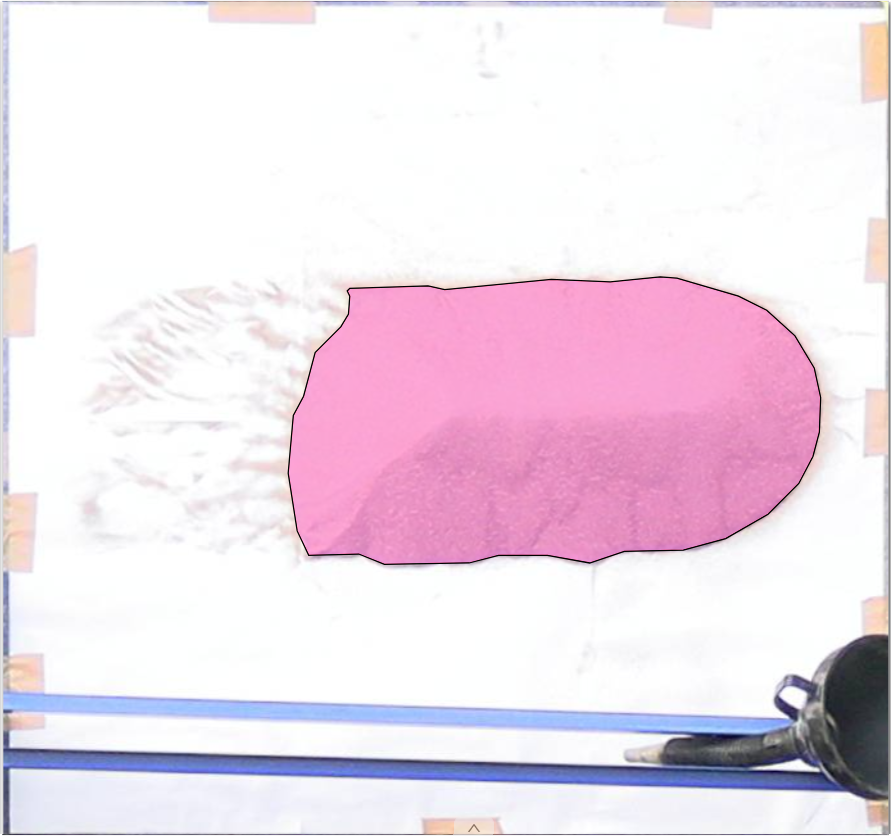}
\includegraphics[width=0.16\columnwidth]{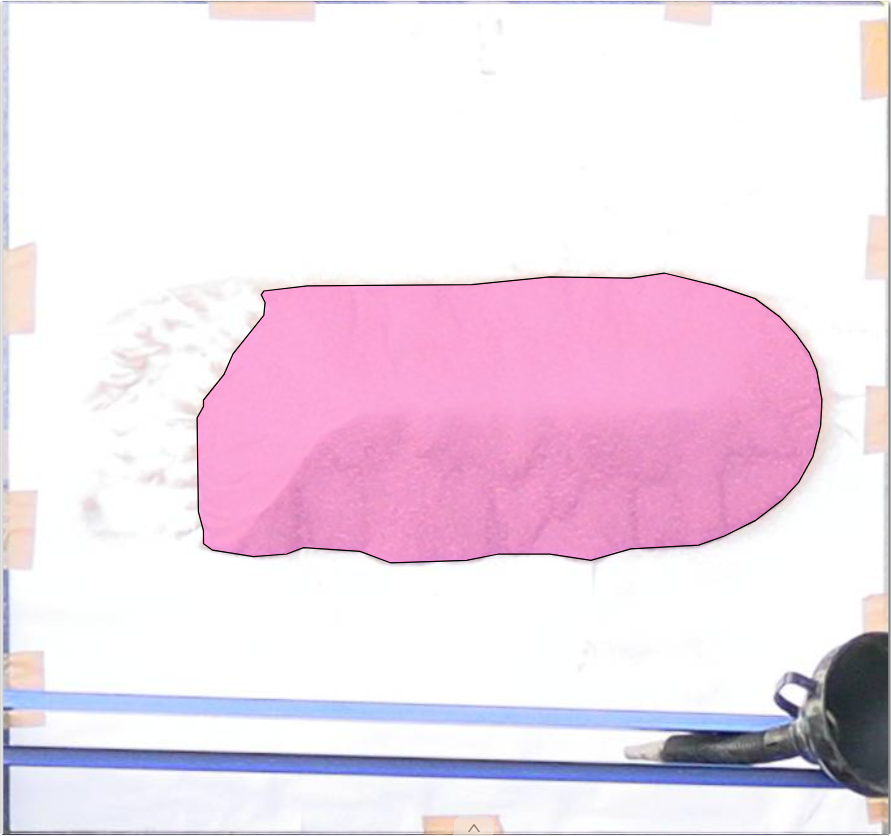}

\includegraphics[width=0.152\textwidth]{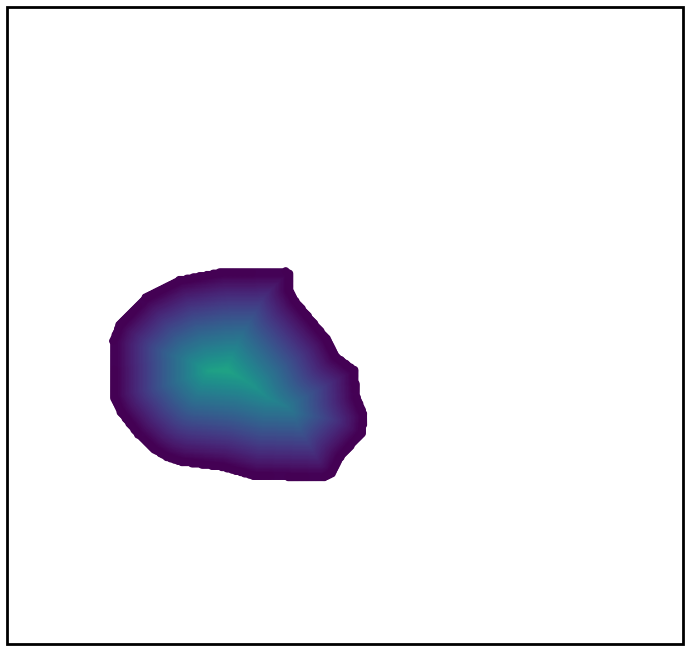}
\includegraphics[width=0.152\textwidth]{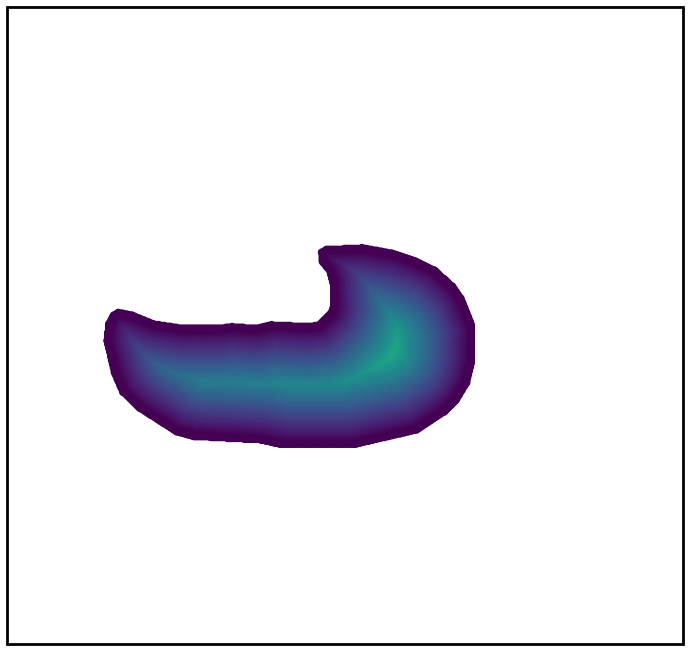}
\includegraphics[width=0.152\textwidth]{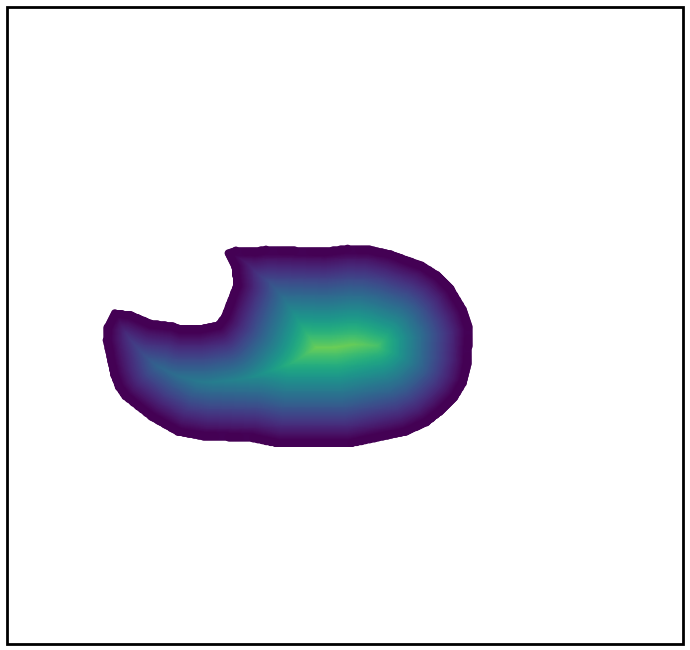}
\includegraphics[width=0.152\textwidth]{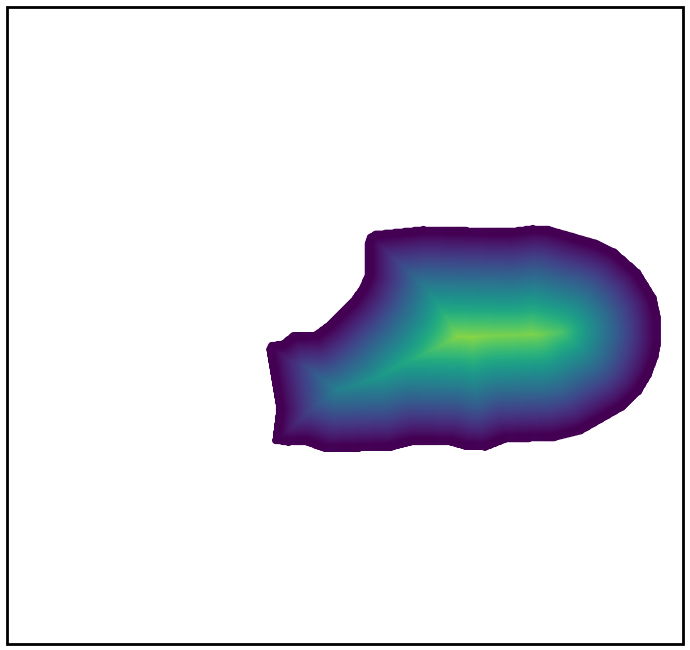}
\includegraphics[width=0.152\textwidth]{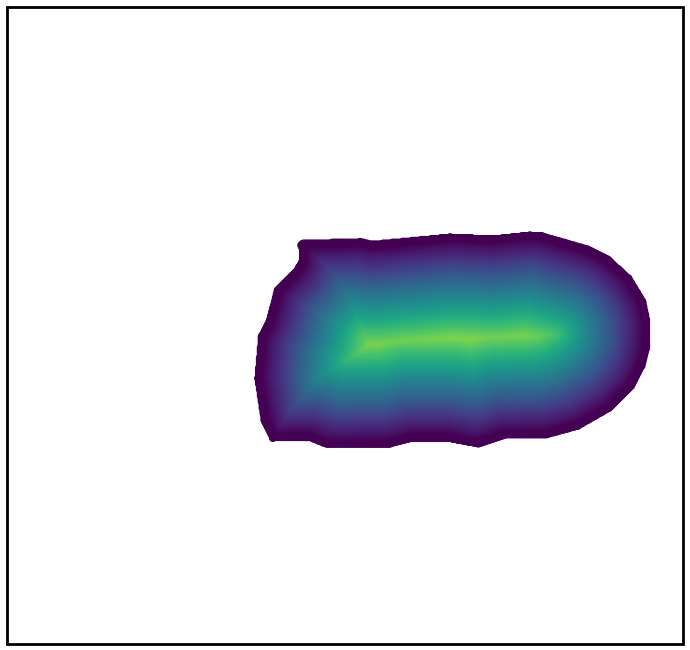}
\includegraphics[width=0.152\textwidth]{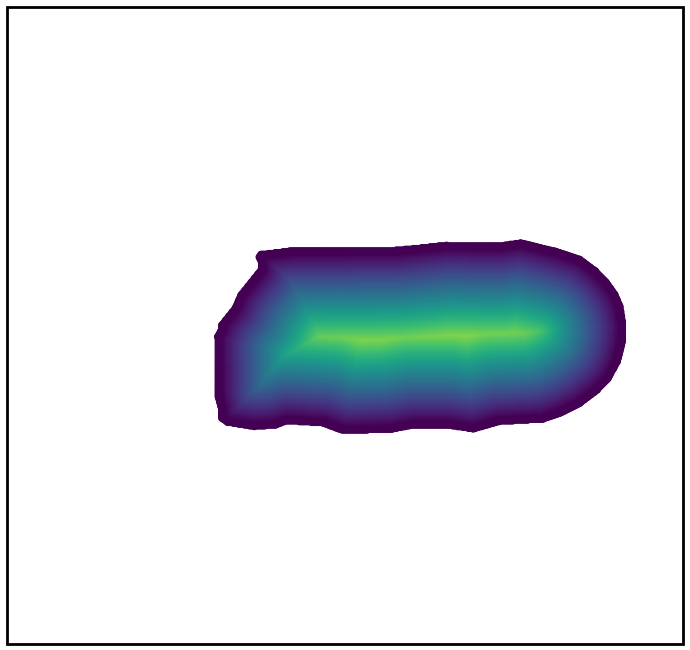}
\includegraphics[width=0.04\textwidth, height = 2.45cm]{colorbar.png}

\caption{Elongated reclaimed (top), segmentation masks (middle) and elongated reclaimed reconstructed (bottom) piles}
\label{fig:elongated_piles_reclaimed}
\end{figure*}

\begin{figure}[h!bt]
    \centering
    \includegraphics[width=0.32\columnwidth]{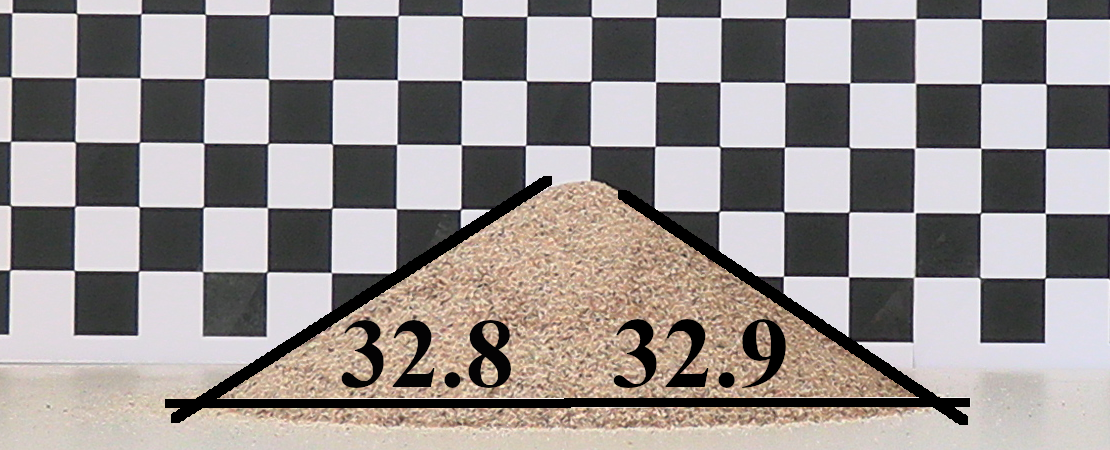}
    \includegraphics[width=0.32\columnwidth]{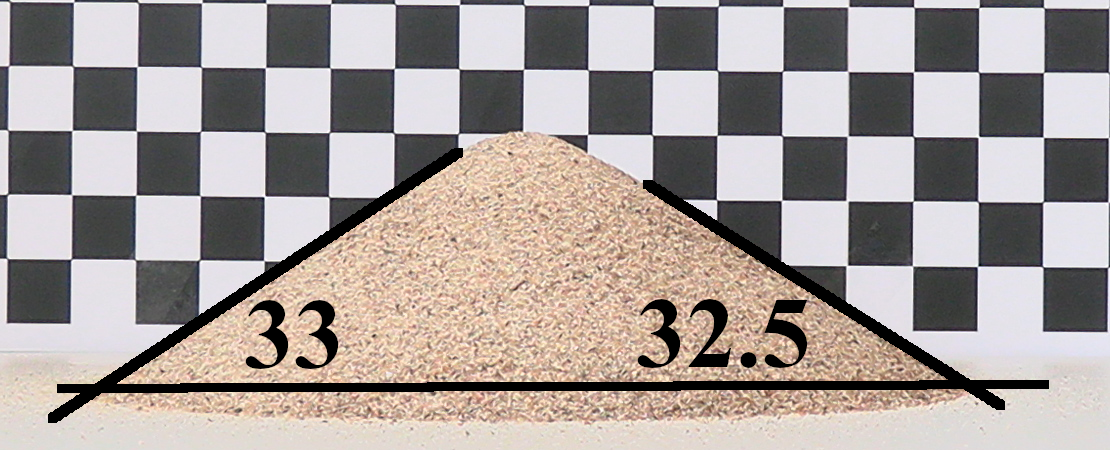}
    \includegraphics[width=0.32\columnwidth]{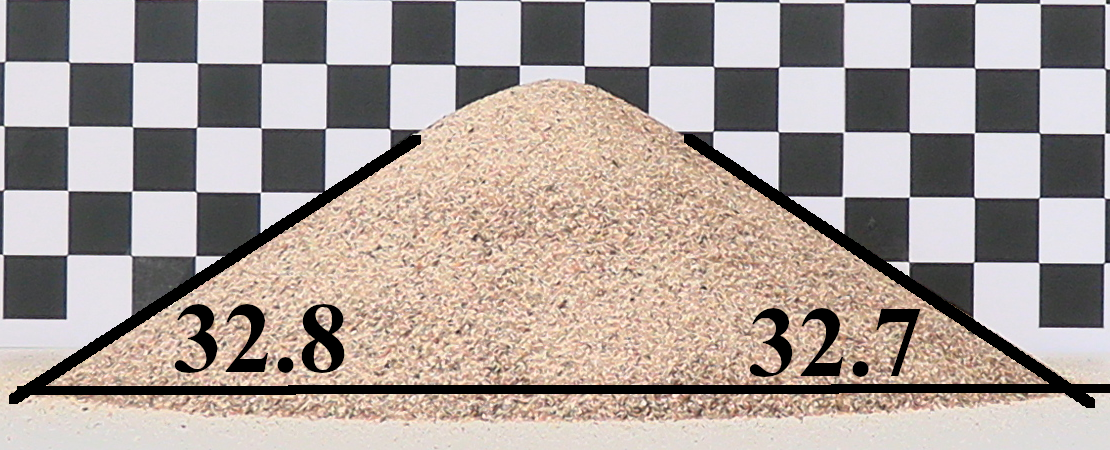}
    \caption{Measuring \emph{angle of repose} from three different piles}
    \label{fig:angle_of_repose}
\end{figure}

\begin{figure}[h!bt]
    \centering
    \includegraphics[origin = c, width=0.8\columnwidth]{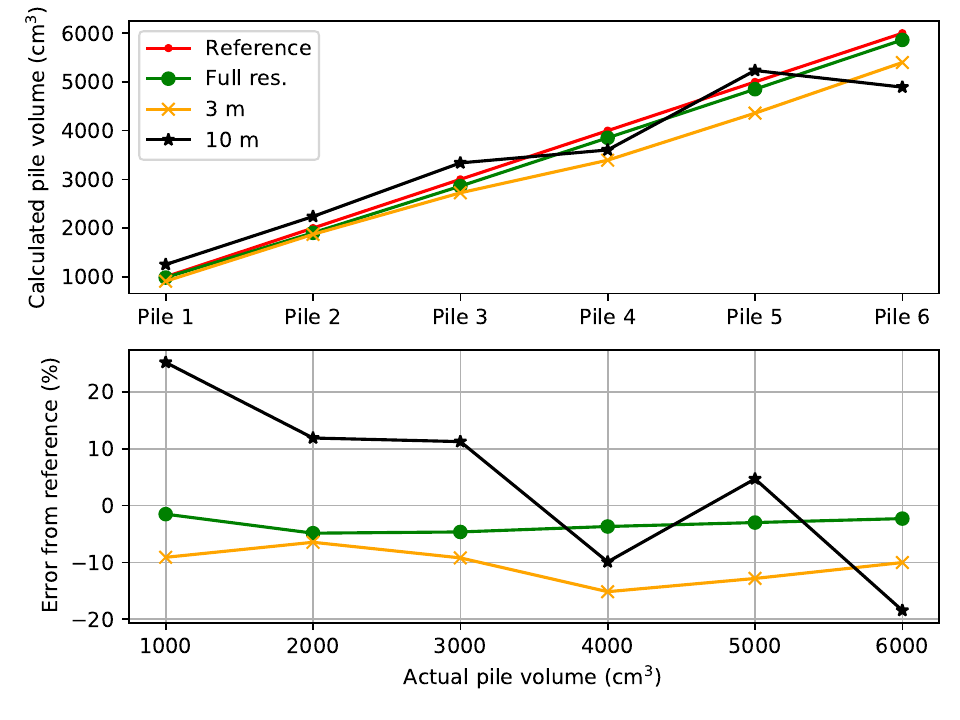}          
    \caption{Volume and error calculation for conical full piles}
    \label{graph:conicalfull}
\end{figure}

\begin{figure}[h!bt]
    \centering
    \includegraphics[origin = c, width=0.8\columnwidth]{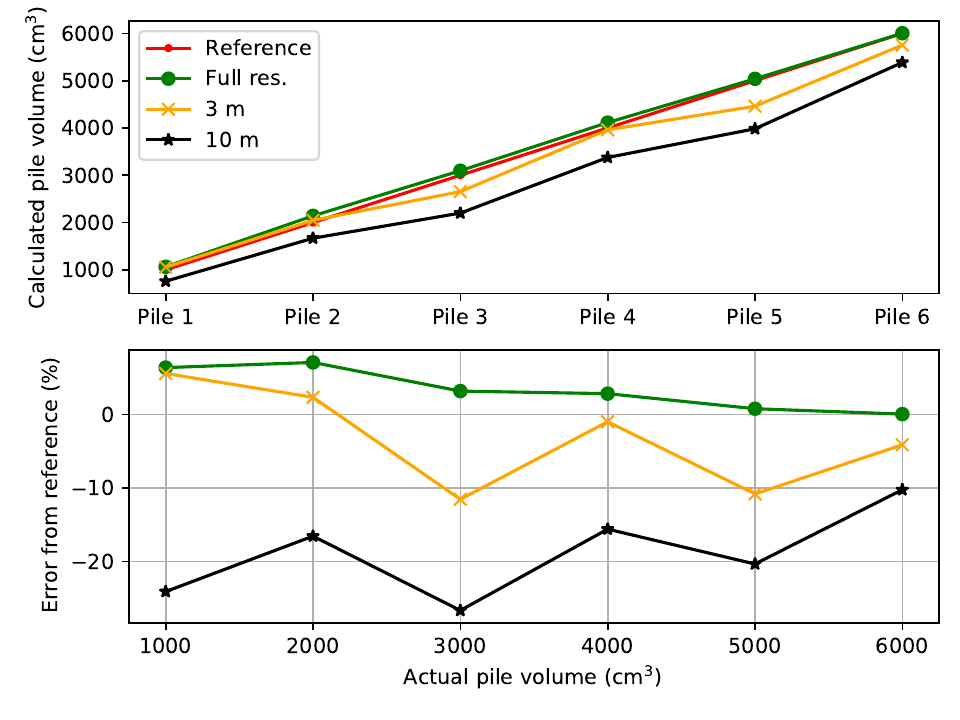}          
    \caption{Volume and error calculation for conical reclaimed piles}
    \label{graph:conicalreclaimed}
\end{figure}

\begin{figure}[h!bt]
    \centering
    \includegraphics[origin = c, width=0.8\columnwidth]{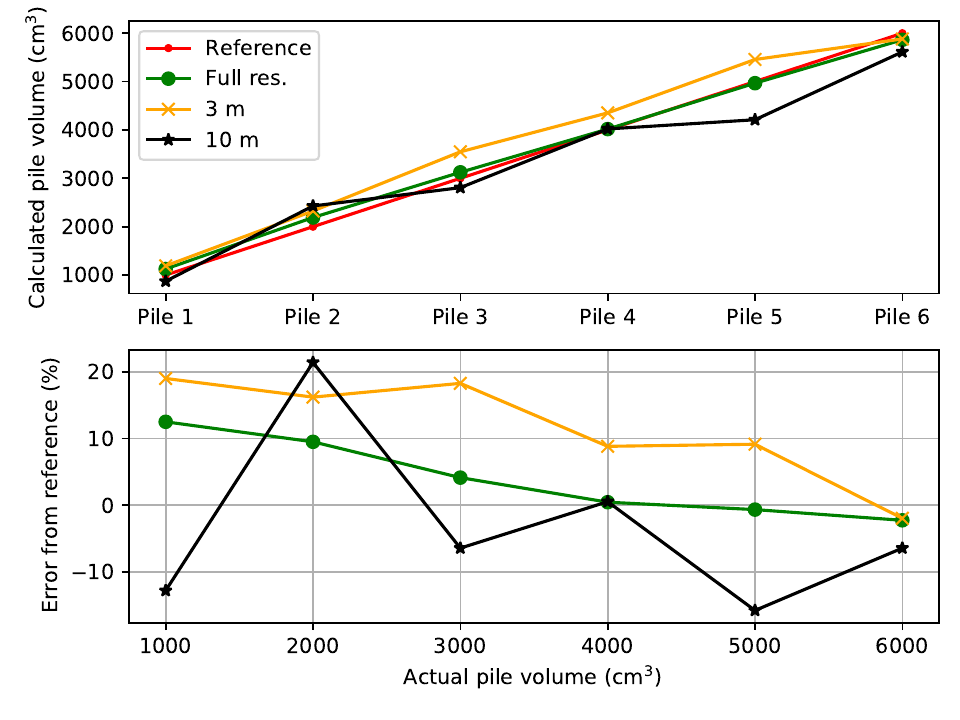}          
    \caption{Volume and error calculation for elongated full piles}
    \label{graph:elongatedfull}
\end{figure}

\begin{figure}[h!bt]
    \centering
    \includegraphics[origin = c, width=0.8\columnwidth]{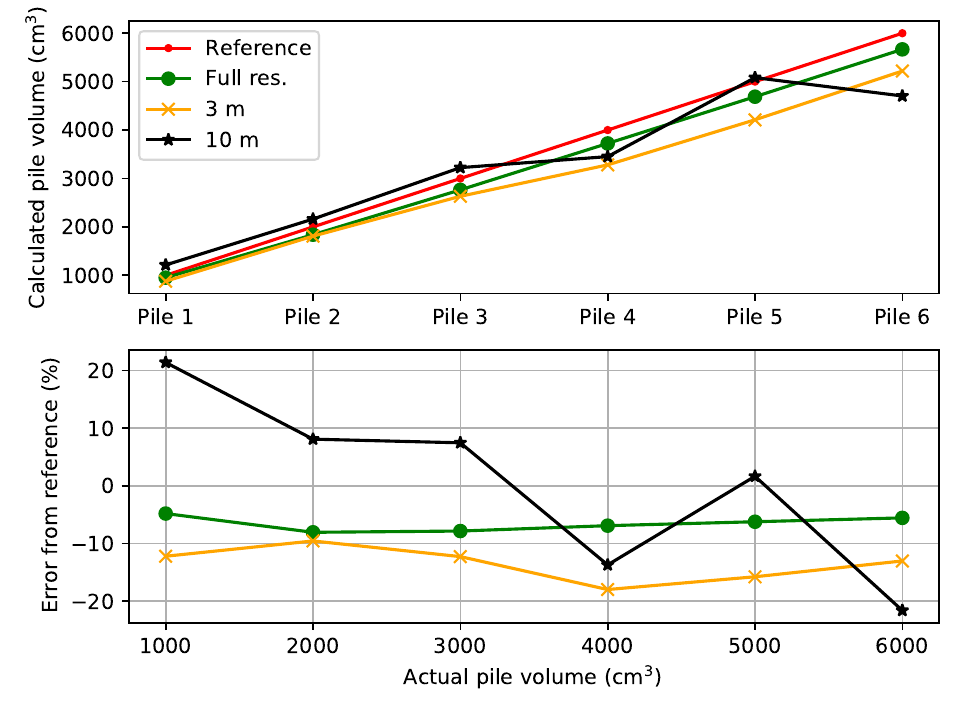}          
    \caption{Volume and error calculation for elongated reclaimed piles}
    \label{graph:elongatedreclaimed}
\end{figure}

Figures~\ref{fig:conical_piles_full}, ~\ref{fig:elongated_piles_full},~\ref{fig:conical_piles_reclaimed},~\ref{fig:elongated_piles_reclaimed} show full conical piles, reclaimed conical piles, full elongated piles and reclaimed elongated piles, respectively. Each of these figures shows piles, segmentation masks, and reconstructed piles using the algorithm. 
 All volume calculations and their corresponding errors are plotted in Figures~\ref{graph:conicalfull}--~\ref{graph:elongatedreclaimed}. \textcolor[rgb]{0.05,0.05,0.05}{To validate the algorithm’s performance, we employ percentage error as the evaluation metric, calculated between estimated and ground-truth volumes.
}The measured length of \SI{84}{cm} covers 650 pixels in the image. This equivalence is used as a conversion factor for converting pixels into real-world lengths for calculating the volume of piles. Figure ~\ref{fig:angle_of_repose} shows the three different piles we used for the angle of repose calculation. We used the mean (\SI{32.78}{\degree}) of the angle values of these piles when calculating the theoretical and experimental volume from the contours of the piles.


\begin{figure}[h!bt]
    \centering
    \includegraphics[width=0.4\columnwidth]{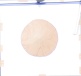}
    \includegraphics[width=0.4\columnwidth]{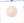}    
    \caption{Downsampled images to \SI{3}{\meter} (left) and \SI{10}{\meter} (right) spatial resolution}
    \label{fig:downsampledimages}
\end{figure}

We tried to simulate three different spatial resolutions of satellite-very high, high, and medium, using the experimental images. For reference, we took examples of PlanetScope images from the Yandi mine, Australia. These images have \SI{3}{m} spatial resolution, and the full piles are approximately \SI{50}{m} wide and \SI{200}{m} long. The first one we try to simulate is very high resolution (approximately \SI{35}{\centi\meter}) using the original experimental images.  Taking the largest elongated full ($190 \times 560$ pixels) pile from our experiment as a reference, we calculate that each of the \SI{560} pixels represents a distance of approximately \SI{35}{\centi\meter} assuming a \SI{200}{\meter} long pile. The other ones are \SI{3}{\meter} and \SI{10}{\meter} spatial resolution using downsampled images. To simulate the low and medium resolution satellite images, we have downsampled the images from the experiment and applied the algorithm to those images. The downsampling factor was determined on the basis of the dimensions of the real-world piles. As the reference images of PlanetScope represent \SI{3}{m} of the real-world distance, these piles are approximately $16 \times 66$ pixels. We downsampled the experimental images by 8.4 times and found that the pile became approximately $23 \times 67$ pixels. With this factor, we mimic the PlanetScope \SI{3}{\meter} spatial resolution images. In addition, to mimic Sentinel 2 images which have spatial resolution of \SI{10}{m}, we have downsampled images by a factor of 28 $(8.4 \times 10 /3)$. 
 A sample each of \SI{3}{\meter} and \SI{10}{\meter} spatial resolution downsampled image is presented in~\ref{fig:downsampledimages}.  Other downsampled images, their segmentation masks, and reconstructed images are not included to avoid overpopulation of images in the paper. These images are provided for the readers as supplementary materials in a separate file.

\begin{figure}[h!bt]
    \centering
    \includegraphics[width=0.4\columnwidth]{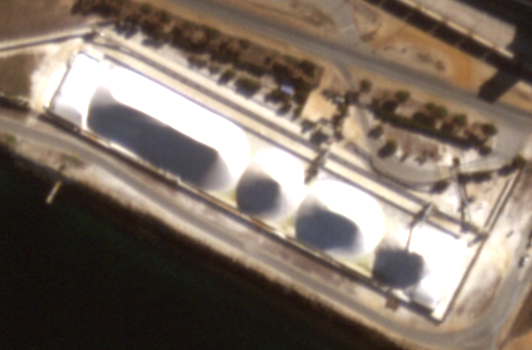}
    \includegraphics[width=0.4\columnwidth]{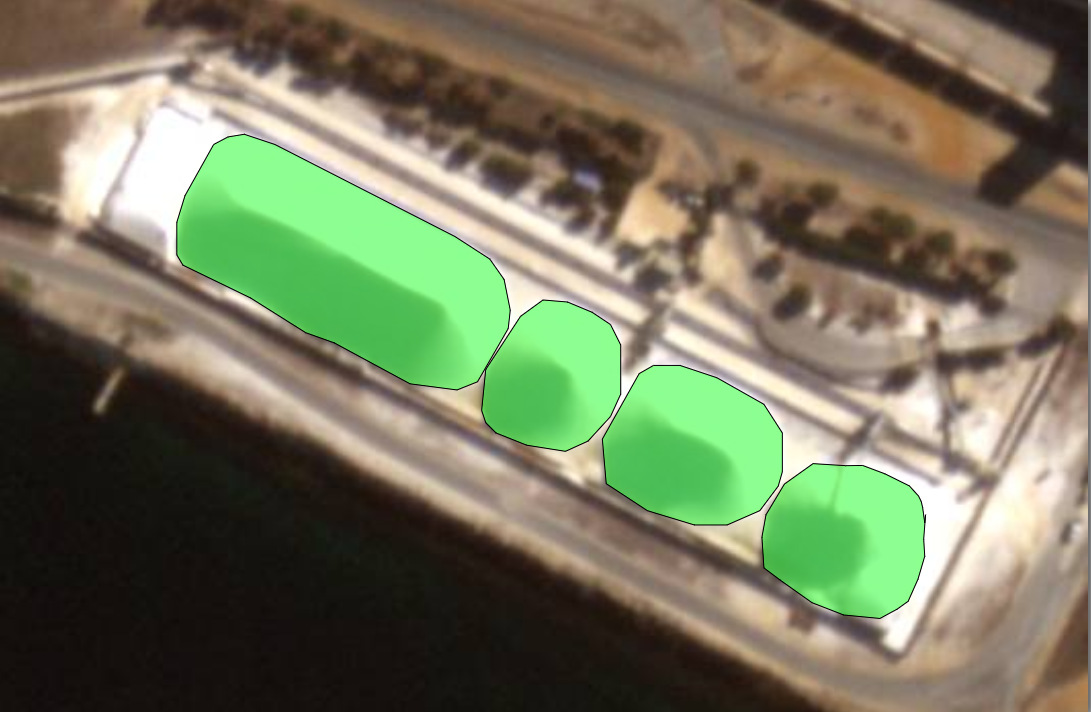}
    \caption{(L to R): Planet's SkySat imagery of \SI{50}{\cm} spatial resolution taken on April 19, 2024 from AustSand's silica sand storage facility ($35^\circ 02' 09'' \, \text{S}, \, 117^\circ 53' 58'' \, \text{E}$) on Princess Royal Drive at the Albany Port, Australia and corresponding segmented image showing the masks of the piles}
    \label{fig:silica_sand_piles}
\end{figure}

Table~\ref{tab:volumeusingsatelliteimage} shows calculation of volume using the algorithm for SkySat image. According to port operations information in Austsand Mining \citep{austsandmining2025}, approximately \SI{40}{\kilo\tonne} of different grades of silica can be stored on site at a time. Since the exact volume of each of the pile on the date of image acquisition is unknown, the maximum recorded volume of silica was used as a reference for comparison. This belief is further strengthened by the site image showing completely full of silica piles. The total tonnage calculated using this method is \SI{45.5}{\kilo\tonne}, which gives an overestimation of 13.7\% compared to the reference value \SI{40}{\kilo\tonne}. This error, among other error sources discussed in Subsection \ref{sources_of_error}, can also be attributed to different angles of repose corresponding to different grades of silica. As we do not know the exact angle of repose for each pile, we have used the same value for all piles.

\begin{table}[!htb]
\centering
\caption{Estimating Weight of silica sand piles. \\$\theta_c = 33.8^\circ$~\citep{anwar2021comparative} and bulk weight density of \SI{1.6}{\tonne\per\m\cubed}~\citep{EuroquarzSilicaSand}}
\begin{tabularx}{\columnwidth}{ccc}\toprule
     Pile. no. & Volume (\unit{\m\cubed}) &   Weight (\unit{\kilo\tonne})\\
     \midrule
     1 & 15156.62  & 24.25 \\
     2 &  3581.5 & 5.7  \\
    3 &  5688.625 & 9.1\\
     4 &  4021.75  & 6.4 \\
     \bottomrule
\end{tabularx}

\label{tab:volumeusingsatelliteimage}
\end{table}

\section{Discussion} 
The objective of this study is to test a new volume estimation algorithm using experimental and satellite images. From the results (Figures \ref{graph:conicalfull}-\ref{graph:elongatedreclaimed}), the Full res. volumes align closely to the Reference volumes for very high full-resolution images for conical full (Figure~\ref{graph:conicalfull}),  reclaimed (Figure~\ref{graph:conicalreclaimed}), elongated full (Figure~\ref{graph:elongatedfull}), and reclaimed (Figure~\ref{graph:elongatedreclaimed}) piles. As the resolution decreases,  the error increases through under- or overestimation as indicated by \SI{3}{\meter} and \SI{10}{\meter} lines. In addition, these results show that the algorithm performs consistently for both full and reclaimed piles, as we do not see any systematic under- or overestimations related to any pile type. Below, we present a discussion on different sources of errors (section~\ref{sources_of_error}), the calculations on downsampled images (section~\ref{validation_on_downsampled_images}), and some limitations (section~\ref{limitations}) of the algorithm. 

\subsection{Sources of Error}
\label{sources_of_error}
\begin{figure}[h!bt]
    \centering
    \includegraphics[width=0.5\columnwidth]{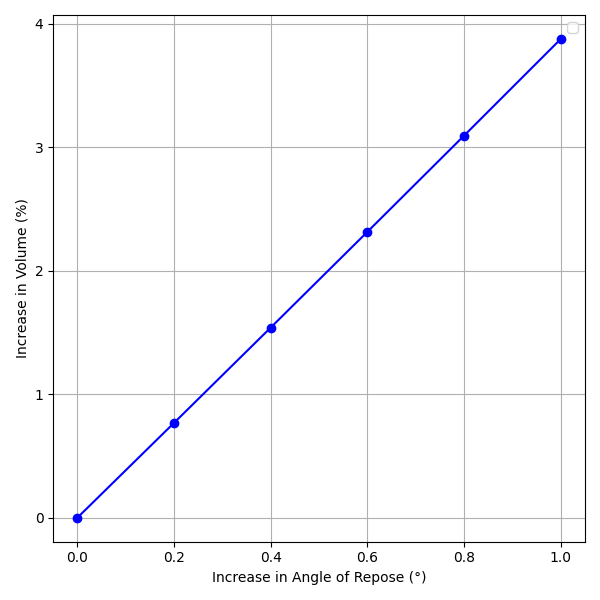}
    \includegraphics[width=0.5\columnwidth]{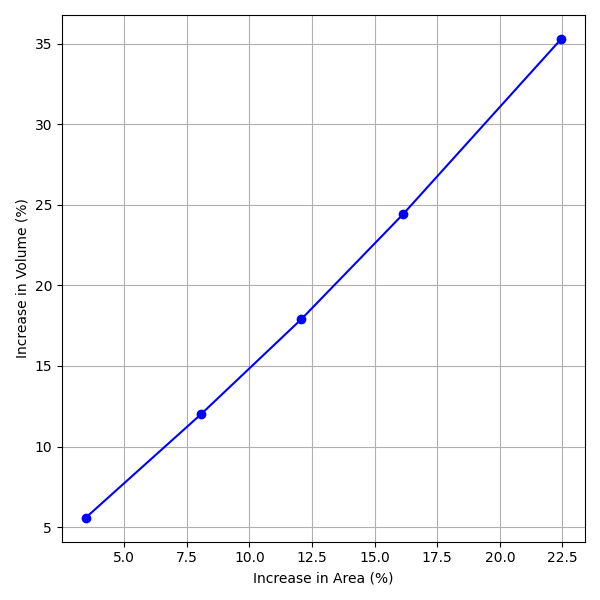}    
    \caption{Volume change with change in angle of repose (top) and volume change with change in contour size (bottom)}
    \label{fig:volumeerror_angle_contour}
\end{figure}

The first major source of error comes from the measurement of the angle of repose as measured in Figure~\ref{fig:angle_of_repose}. If the angle of repose is less than the true value, it causes a consistent underestimation in the volume calculation. Similarly, volume is overestimated if the measured angle of repose is higher than its true value, as shown in Figure~\ref{fig:volumeerror_angle_contour}. Angle of repose and contour are coupled during pile formation, making the base smaller for steeper angles and wider base for less steep angles. However, we decouple them for analyzing their separate effect on volume calculation. This helps us to see how volume changes with under- or overestimation of the angle of repose, keeping the contour fixed and vice versa. The plots in Figure~\ref{fig:volumeerror_angle_contour} are made using the \SI{1000}{cm^3} conical full pile. The figure on the left shows how much volume increases with an increase in the angle of repose. We increased the angle of repose in small steps and calculated the volume. With an increase of \SI{1}{\degree} in the angle of repose, the volume overestimation is around \SI{4}{\percent} compared to the base volume calculated using the true angle of repose. Volume increment (Vol\_Inc) with respect to angle of repose (AoR) is governed by the equation 
\begin{equation}
    Vol\_Inc = 3.8781 \cdot AoR - 0.0059
    \label{eq:vol_angle_of_repose}
\end{equation}
Equation~\ref{eq:vol_angle_of_repose} shows that volume increases linearly with the increase in AoR when the contour remains unchanged.

As \SI{10}{\meter} resolution images give inconsistent under- and overestimation, we refer to the errors of Full res. and \SI{3}{\meter} images for analysis of the errors in volume estimation. The volume is clearly underestimated in conical full and elongated reclaimed piles shown in Figures~\ref{graph:conicalfull} and ~\ref{graph:elongatedreclaimed}, respectively. For conical reclaimed piles (Figure ~\ref{graph:conicalreclaimed}), there is a slight overestimation for Full res., while most of the volume calculation is underestimated for \SI{3}{\meter} images. For elongated full piles (Figure \ref{graph:elongatedfull}), there is some overestimation and some underestimation for \SI{3}{\meter} images, while mostly it is underestimated for Full res. This shows that volume has been mostly underestimated, indicating a possible undervaluation of the angle of repose while measuring it. These small experimental piles may have slight asymmetries due to small amount of pouring, causing the pile to accumulate more towards a certain direction. We used images from one direction to estimate the angle of repose. More images from multiple directions and averaging the values of angles could give us a more accurate estimation.

 The method is also prone to errors caused by the contour of the pile. As the method is based on building a 3D model using the contour, a small error in the contour can be magnified in the volume calculation process. Figure~\ref{fig:volumeerror_angle_contour} on the right shows the error in volume that transfers from the error in the contour area. To plot this, we manually changed the segmentation masks by stretching the points of the contour after the masks were formed from the detection of the piles using SAM 2, and calculated the change in volume. The plots show that the volume error is always higher than the error in the contour area. The equation that governs the increase in volume with respect to the increase in the area of contour (AoC) is given by 
 \begin{equation}
     Vol\_Inc = 1.5661 \cdot AoC - 0.4326
    \label{eq:vol_area_of_contour}
 \end{equation} Keeping the angle of repose constant, this equation shows a linear increase in volume with respect to an increase in AoC. 
 We can see the segmentation masks of the piles in the images presented in the Figures~\ref{fig:conical_piles_full}-~\ref{fig:elongated_piles_reclaimed}. These contours are mostly accurate with small inaccuracies along the boundaries. 
 
 However, the contours may be inaccurate in satellite images as the background and piles share the same material, resulting in errors in the volume estimation. In  Figure~\ref{fig:silica_sand_piles}, we can clearly notice small irregularities in the detected contours. In addition, the boundary is clear on the shadowed side, but not so clear on the other sides. These inaccuracies in the contours are transferred to the volume calculation in the same way as the errors calculated in Figure~\ref{fig:volumeerror_angle_contour}.

\subsection{Validation with Downsampled Images}
\label{validation_on_downsampled_images}
For the downsampled images, the segmentation algorithm struggles to detect the correct boundaries, making the segmentation masks less smooth and accurate (see the supplementary materials for the figures).  The reconstructed piles are pixilated because fewer pixels are used to represent the same piles. With images mimicking the spatial resolution of the \SI{3}{\m} satellite, the error increased compared to the original resolution. It is $<20\%$ for both conical and elongated full and reclaimed piles, as indicated by \SI{3}{\meter} lines in the lower figures in the plots~\ref{graph:conicalfull}-~\ref{graph:elongatedreclaimed}. With a further decrease in the spatial resolution to \SI{10}{m}, the error increases up to $<30\%$ as well as being more random, shown by \SI{10}{\meter} lines in the same figures. The randomness is apparent in the jumps from overestimation to underestimation. As the only thing we have changed here is the resolution of images, as expected, the lower resolutions result in higher uncertainty in volume calculation.
The above analysis shows that using \SI{3}{\meter} or higher resolution images produces better segmentation masks and low errors in volume estimations.

\subsection{Limitations}
\label{limitations}
In a practical application of this method to different types of piles, we expect some of the piles to cause the method to overestimate the amount of stored material. Especially flat-topped piles will be wrongly estimated, as the method used here assumes that the piles are loaded to their maximum height. Flat-topped piles can be the result of several limitations, like the maximum height of the unloader, load limit on the storage yard, or others. Although in large ports, we generally find fully loaded and their reclaimed piles, it is more common in smaller ports to find flat-top and their reclaimed piles. The algorithm does not work for these types of piles.

We have also shown the applicability of the method to satellite images. In our experiment, we used a right-angle top-down camera from a low height, which does not need us to do atmospheric corrections on the images. If we use satellite images for volume calculation, we may need an atmospheric correction step before applying this algorithm. The background of the experimental images is plain white, which is not the case in satellite images. Real-world pile images from satellite, drone, or aircraft have the same pile material as their background. These two factors are the main considerations that we need to be aware of if we apply this algorithm to real-world images. However, provided the detected boundary is accurate, the algorithm will still give accurate volume estimation even in satellite images, as shown by the accuracy of estimation using the original experiment images.

\textcolor[rgb]{0.05,0.05,0.05}{The current algorithm assumes a constant angle of repose for all kinds of piles and does not explicitly account for variations arising from different environmental factors such as moisture or wind. Other factors that need to be considered are properties of the material such as grain size distribution and density \citep{ramadan2024experimental, azzam2018utilization,azzam2025geotechnical, sallam2024effects}. While incorporating these parameters could improve the accuracy of volume estimation, the present study focuses on evaluating the algorithm under controlled experimental conditions. Future work could extend the algorithm to include environmental influences by integrating these additional parameters.  }\\
\textcolor[rgb]{0.05,0.05,0.05}{Another limitation comes from the algorithm’s assumption that piles are formed uniformly according to the angle of repose in all directions, resulting in idealized symmetric shapes. In practice, however, real-world piles may deviate from this assumption due to irregular deposition, particle segregation or uneven surface. Such deviations can introduce systematic under- or over-estimations in the calculated volumes. Future work could focus on going beyond the uniformity assumption and extending the algorithm to incorporate these factors, thereby improving its applicability to complex real-world piles.}

\section{Conclusion}

This paper successfully confirms the accuracy of the volume calculation algorithm. The maximum error calculated is around 25\% for images of \SI{10}{\meter} resolution. An increase in the resolution consistently decreases the error. The algorithm provides a much simpler alternative to estimating the volume of the cargo pile using a single satellite image. The accurate and timely estimation of volume helps enhance logistics management, optimizes routing and scheduling of ships, and support decision making in port operations. Validation through the use of experimental and satellite images (SkySat) provides enough evidence for the accuracy of the volume calculation algorithm based on angle of repose.


\section*{Acknowledgement}

This work is supported by the Norwegian Research Council [contract \#326609].

\bibliographystyle{model2-names}

\bibliography{cas-refs}

\end{document}